\documentclass[preprint]{aastex}
\usepackage{hyperref}
\usepackage{rotating}
\pdfoutput=1

\begin{document}

\title{Infall/Expansion Velocities in the Low-Mass Dense Cores L492, L694-2, and L1521F: Dependence on Position and Molecular Tracer}

\author{Jared Keown\altaffilmark{1,2,3}, Scott Schnee\altaffilmark{1}, Tyler L Bourke\altaffilmark{4,5}, James Di Francesco\altaffilmark{6,3}, Rachel Friesen\altaffilmark{7}, Paola Caselli\altaffilmark{8}, Philip Myers\altaffilmark{4}, Gerard Williger\altaffilmark{2,9,10}, Mario Tafalla\altaffilmark{11}}

\email{jkeown@uvic.ca}


\altaffiltext{1}{National Radio Astronomy Observatory, 520 Edgemont Rd, Charlottesville, VA 22903, USA}

\altaffiltext{2}{Department of Physics and Astronomy, University of Louisville, 102 Natural Science Building, Louisville, KY 40292, USA} 

\altaffiltext{3}{Department of Physics and Astronomy, University of Victoria, Victoria, BC, V8P 5C2, Canada} 

\altaffiltext{4}{Harvard-Smithsonian Center for Astrophysics, 60 Garden Street, Cambridge, MA 02138, USA }

\altaffiltext{5}{SKA Organisation, Jodrell Bank Observatory, Lower Withington, Macclesfield, Cheshire SK11 9DL, UK} 

\altaffiltext{6}{NRC Herzberg Astronomy and Astrophysics, 5071 West Saanich Road, Victoria, BC, V9E 2E7, Canada}

\altaffiltext{7}{Dunlap Institute for Astronomy and Astrophysics, University of Toronto, 50 St. George St., Toronto, ON, M5S 3H4, Canada}

\altaffiltext{8}{Max-Planck-Institute for Extraterrestrial Physics (MPE), Giessenbachstr. 1, D-85748 Garching, Germany}

\altaffiltext{9}{Jeremiah Horrocks Institute, University of Central Lancashire, Preston PR1 2HE, UK}

\altaffiltext{10}{Institute for Astrophysics and Computational Sciences, Catholic University of America, Washington, DC 20064, USA}

\altaffiltext{11}{IGN, Observatorio Astron\'omico Nacional, Alfonso XIII 3, 28014, Madrid, Spain}


\begin{abstract}

Although surveys of infall motions in dense cores have been carried out for years, few surveys have focused on mapping infall across cores using multiple spectral line observations. To fill this gap, we present IRAM 30-m Telescope maps of N$_2$H$^+$(1-0), DCO$^+$(2-1), DCO$^+$(3-2), and HCO$^+$(3-2) emission towards two prestellar cores (L492 and L694-2) and one protostellar core (L1521F).  We find that the measured infall velocity varies with position across each core and choice of molecular line, likely as a result of radial variations in core chemistry and dynamics.  Line-of-sight infall speeds estimated from DCO$^+$(2-1) line profiles can decrease by 40-50 m s$^{-1}$ when observing at a radial offset $\geq$0.04 pc from the core's dust continuum emission peak.  Median infall speeds calculated from all observed positions across a core can also vary by as much as 65 m s$^{-1}$ depending on the transition.  These results show that while single-pointing, single-transition surveys of core infall velocities may be good indicators of whether a core is either contracting or expanding, the magnitude of the velocities they measure are significantly impacted by the choice of molecular line, proximity to the core center, and core evolutionary state.     

\end{abstract}

\section{Introduction}

Molecular gas and dust cores with central densities of 10$^5$ cm$^{-3}$, temperatures of $\sim$ 10~K, and diameters of $\sim$ 0.1~pc serve as the cocoons out of which stars are born \citep{andre_2013}.  An evolutionary stage classification scheme for cores based upon the presence or lack of a forming protostar has been developed in recent years.  Cores with observable submillimeter continuum emission, but without a detectable infrared source, are generally characterized as ``starless" because they have not yet formed a protostar \citep{1994MNRAS.268..276W}.  Starless cores undergo collapse when the inward force of gravity overcomes the outward push of the internal pressure of the system, at which point they are termed ``prestellar.''  Prestellar cores are thus gravitationally contracting, with material moving toward their centers \citep[e.g.,][]{1999ApJ...526..788L, 2005ApJ...619..379C, Keto_2015}.  The less centrally concentrated starless cores also show expansion and oscillation motions \citep[e.g.,][]{0004-637X-586-1-286, Tafalla_2004}.  This phenomenon is commonly observed \citep[e.g.,][]{Redman_2006, Aguti_2007, 2007ApJ...664..928S, Shadi_2014} and is thought to be caused by oscillatory motions of gas in the outer layers of the core \citep{Broderick_2010}.  Observations by \cite{2011ApJ...734...60L} also suggest that oscillating cores serve as an evolutionary bridge between static and collapsing cores.

On the chemical level, radial position within the core plays a major factor in the relative abundances of many molecules in the gas phase.  Certain carbon-based molecules, such as CO, freeze onto the surface of dust grains at temperatures around 10~K and densities above 10$^4$ cm$^{-3}$ \citep{1999ApJ...523L.165C}.  As a result, gas phase CO, along with other carbon-bearing molecules, can be significantly depleted toward core centers \citep{Tafalla_2002}.  A side effect of CO depletion is the confinement of certain molecules to the centers of cores. For example, nitrogen-bearing molecules appear to survive in the gas phase at higher densities and lower temperatures than carbon-based molecules.  Here, the individual nitrogen atom, which may have a lower binding energy than CO, may be transformed into N$_2$ via slower neutral-neutral reactions \citep[see, e.g.,][]{Flower_2006, Hily-Blant_2010}.  Once N$_2$ is produced, other easily detected N-bearing species (such as N$_2$H$^+$ and NH$_3$) are produced.  N$_2$H$^+$ is found most abundantly in the cold, high density, inner regions of cores where its main reactants, e.g. CO, are depleted due to freeze-out \citep{2007prpl.conf...17D}.  For a similar reason, deuterated molecules are also found most abundantly in core centers.  CO and molecules containing CO such as HCO$^+$ slow-down the production of deuterium-enriched particles in outer core regions by reacting with H$_2$D$^+$, one of the main precursors to deuterium fraction in molecules such as HCO$^+$, N$_2$H$^+$, and NH$_3$ \cite[e.g.,][]{Aikawa_2012}.  Since CO is depleted towards the central regions, however, deuterated molecules with easily detectable rotational lines, such as N$_2$D$^+$, can be formed \citep{2005ApJ...619..379C}.  Studies have also linked deuterium enrichment to physical evolution, with a higher fraction of deuterated molecules corresponding to a more dynamically evolved core \citep{2005ApJ...619..379C, schnee_2013}.

Rotational lines from molecules are often used to trace the structure, kinematics, and chemistry of dense cores \citep[e.g.,][and references therein]{2007ARA&A..45..339B}.  Based upon the shape of the observed emission line profile, one can determine various core properties such as the infall or expansion velocity of the gas.  Doppler shifts induce asymmetries in the spectra observed from either contracting or expanding cores.  When observing optically thick emission from a contracting core (infall), an asymmetrically blue, double-peaked line profile with a blue peak brighter than a red peak can be observed.  The weaker red peak is caused in part by: 1) \textit{obscuration} of the higher excitation, redshifted material near the center of the core by lower excitation material located on the outskirts and 2) \textit{emission} from the outer, lower excitation material.  Observations of optically thick lines can also result in a self-absorption dip near the velocity centroid of the core.  For an expanding core, observations of optically thick emission result in line profiles that are asymmetrically red with a red peak brighter than a blue peak.  In this scenario, the central, high excitation material being obscured is on the blueshifted end of the spectrum due to the outward motions of the core.  


Prestellar and protostellar core contraction have been modelled extensively to produce a wide range of solutions.  Theoretical models by \cite{1977ApJ...214..488S} predict that the collapse of dense cores starts on the inside and moves out, with the fastest motions at the center.  Other solutions to the collapse problem suggest an outside-in process in which small disturbances initiate contraction on the outer layers of the core which propagate to the central regions \citep{Larson_1969, Penston_1969}.  Observations have attempted to test these models and found that infall is indeed spatially extended across the highest column density regions of prestellar cores \citep{1998ApJ...504..900T, 2001ApJS..136..703L}.  Recent work based on the detection of the ground state ortho-H$_2$O water line toward the prestellar core L1544 with Herschel, as well as other high density tracers observed with other observatories, has shown that quasi-static contraction is in better agreement with observations compared to the inside-out and Larson-Penston models \citep{Keto_2015}.  There have not been sufficient observational data, however, to confirm whether or not the \textit{speed} at which a core is contracting varies with distance from the center, with the exception of the prestellar core L1544.  Core geometrical orientation effects undoubtedly play a role in all infall surveys.  Since the emission we observe is only the component along our line of sight to each core, we only receive a projected fraction of the full infall or expansion speed vector.  If the core is taken to be spherical and collapsing, these line of sight effects alone imply decreasing speeds with distance from the center.  In that case, the full infall component would be seen towards the sphere's absolute center since that line of sight is aligned with the collapse direction, while the slowest speeds would be on the edges where our line of sight is at an approximately 90 degree angle from the collapse direction.  Hence, infall/expansion velocity estimates are likely lower limits.  Furthermore, cores have been found to vary widely in shape, ranging from elongated and filamentary structures that resemble the filamentary clouds in which they reside \citep{Hartmann_2002} to prolate/oblate triaxial spheroids \citep[e.g.,][]{Myers_1991, Ryden_1996, Jones_2001, Tassis_2007, Lomax_2013} and compact, round spheres, complicating interpretations. 

Despite an abundance of surveys investigating infall motions \citep[][etc.]{1999ApJ...526..788L, 2004ApJS..153..523L, 2007ApJ...664..928S, schnee_2013}, few have been awarded the time required to map infall across cores using multiple spectral line observations. \cite{0067-0049-136-2-703} mapped 53 cores in CS(2-1), N$_2$H$^+$(1-0), and C$^{18}$O(1-0), but  since they averaged spectra within the half-maximum contour of their N$_2$H$^+$(1-0) intensity maps, they were unable to comment on the relationship between infall and position within a given core.  To our knowledge, only two previous studies have mapped starless cores in detail to determine infall as a function of position \citep{1999ApJ...513L..61W, 2006ApJ...636..952W}.  Hence, two fundamental questions are still relatively uncertain: (1) Do core infall speeds have a dependency on the position observed within the core? (2) Do core infall speeds have a dependency on the chosen molecular tracer?  Answering these questions will help determine the biases involved in the single-pointing and single-tracer studies conducted over the past two decades.  To determine the rate at which a core is either contracting or expanding, we must find out if it is sufficient to observe a single position within the core using a single transition?  Or is it necessary to map across cores in multiple tracers?   

Based on our current understanding of the dynamics and chemistry of cores, it seems likely that infall velocity is dependent upon position inside the core.  Infall speeds are theorized to decrease with distance from the core center under certain models of prestellar core contraction \cite[e.g.,][]{1977ApJ...214..488S, Basu_1994, Ciolek_1995}, which implies a spatial dependency.  3D to 2D projection effects can also lead to measured infall speed gradients across a core depending on its particular geometry.  Chemically, we also know that molecular abundances vary with position inside a core.  Therefore, one may predict that the infall velocity measured from the spectra of different tracer molecules should show variations as well.  To test this hypothesis, we have obtained IRAM 30-m Telescope maps of N$_2$H$^+$(1-0), DCO$^+$(2-1), DCO$^+$(3-2), and HCO$^+$(3-2) emission towards two prestellar cores (L492 and L694-2) and one protostellar core (L1521F).  We compare these maps for each core to determine if infall shows positional variations and whether or not the chosen molecular tracer has an impact on infall velocity magnitudes.

This paper will describe the observations used for our survey in \S~2, outline the techniques and models used to analyze the data in \S~3, discuss possible interpretations and causes of our velocity measurements in \S~4, articulate goals of future studies in \S~5, and summarize this study in \S~6.

\section{Observations}

The three dense cores selected for this survey have been well studied over the past several years.  L492 and L694-2 have both been classified as prestellar due to the absence of a detectable young stellar object \citep{1999ApJ...526..788L,0004-637X-598-2-1112}.  Although L1521F was originally thought to be prestellar, it has recently been found to be protostellar with a confirmed bipolar outflow originating from an embedded VeLLO \citep{1538-4357-649-1-L37, 0004-637X-774-1-20}.  All three cores have been found to have signatures of infall asymmetry in previous surveys \citep[e.g.,][]{2005ApJ...619..379C, 2007ApJ...664..928S, 2011ApJ...734...60L, schnee_2013}. High-resolution spectral line emission maps have also been observed toward L694-2 \citep{2006ApJ...636..952W} which have shown a radial gradient of infall speeds across the core with decreasing speeds as distance from the center increases.  The \cite{2006ApJ...636..952W} study was limited, however, by the fact that it only used a single molecular tracer (N$_2$H$^+$(1-0)).  Table 1 outlines the physical characteristics of the three targets and includes their infall velocity measurements from the \cite{schnee_2013} single-pointing HCO$^+$(3-2) survey.


The critical density of a molecular transition from an upper level \textit{u} to a lower level \textit{l}, defined as n$_{cr}$(u-l) = A$_{ul}$ / $\gamma_{ul}$, where A$_{ul}$ is the Einstein A-coefficient and $\gamma_{ul}$ is the collisional rate coefficient, provides a rough guideline for the gas density at which observed emission likely originates \citep{doi:10.1146/annurev.astro.37.1.311}.  It has been known for some time, however, that n$_{cr}$ fails to account for effects such as radiative trapping and multilevel excitation \citep[see][]{ doi:10.1146/annurev.astro.37.1.311, Shirley_2015}, which tend to lower the effective density required to detect a line.  Nevertheless, when comparing the relative excitation conditions traced by several transitions, n$_{cr}$ remains a benchmark upon which tracers can be compared.  The four transitions used for our observations were chosen because they have similar values of n$_{cr}$.  Table 2 lists the values of n$_{cr}$ as calculated based on data from the Leiden Atomic and Molecular Database \citep{Schoier_2005} for the four transitions observed here.  The critical densities of the transitions vary by only an order of magnitude at most, which is a small difference considering the fact that densities differ by a few orders of magnitude across a core.  Due to the similar excitation conditions traced by these transitions, they become excellent tracers for investigating any bias involved with choosing an infall tracer solely on its n$_{cr}$ value.  Differences in the chemistry of these tracers, however, may lead to variations in the layers of the core where they are most abundant. Since we are observing N-, C-, and D-bearing molecules, factors such as freeze-out and depletion inevitably play a role in defining the core regions traced by each molecule.  


Our data were obtained at the IRAM 30-m single-dish Telescope in Pico de Veleta, Spain from July-December 2002.  Each core was observed in frequency-switching mode.  Frequency windows were centered on N$_2$H$^+$(1-0) at 93.174 GHz,  DCO$^+$(2-1) at 144.077 GHz, DCO$^+$(3-2) at 216.113 GHz, and HCO$^+$(3-2) at 267.558 GHz.  Beam widths were 27.0$\arcsec$, 17.5$\arcsec$, ~11.6$\arcsec$, and ~9.4$\arcsec$ FWHM for each transition, respectively.  The observed frequencies correspond to a range of emission wavelengths of 3.25 - 1.10 millimeters.  Spectral resolutions were within the range of 0.020 - 0.054 km~s$^{-1}$ depending on the transition, while 1 sigma rms antenna temperature sensitivities were about 20 mK.  (See Table 2 for a summary of these observational characteristics.)  Spectra were obtained at 72 different locations across L694-2, separated by increments of 10$\arcsec$ and 20$\arcsec$.  For L492, a 42 point columned pattern with equal separations of 20$\arcsec$ was used for N$_2$H$^+$(1-0) and DCO$^+$(2-1) while the DCO$^+$(3-2) and HCO$^+$(3-2) maps contain 31 and 16 pointings, respectively.  Lastly, a 42 point columned pattern with separations of 20$\arcsec$ was also adopted for L1521F, except for HCO$^+$(3-2) for which only 10 pointings were observed.  Figure 1 shows the MAMBO 1.2 mm dust continuum emission from \cite{kauffmann_2008} for L492 and the SCUBA 850~$\mu$m dust continuum emission from L694-2 and L1521F, each overlaid with circles that represent the points at which spectra were measured in each core.  (The SCUBA Legacy catalogue of \cite{0067-0049-175-1-277} does not contain L492.)  After acquisition, the data were reduced using standard procedures for IRAM 30-m data using the GILDAS\footnote{http://www.iram.fr/IRAMFR/GILDAS} package.  Frequency axes were converted to velocities and written to text files using GILDAS CLASS routines. 


\section{Analysis}
Figure 2a displays the DCO$^+$(2-1) spectra observed toward L492.  Similar figures for the other three transitions and two additional cores appear in the online version of the Journal as Figures 2b-2d, 3, and 4.  Although most spectra of all transitions show blue asymmetries, we also observe asymmetrically red spectra at multiple pointings in each core.  Figure 5 shows sample spectra from each core with both blue and red asymmetries.  We estimated the noise in a given spectrum using the standard deviation of the flux on the parts of the spectrum where no clear line emission was recorded.  These values were used along with peak brightness temperature to determine the signal-to-noise ratio (SNR) of each spectrum.  SNR values varied depending on the core, transition, and pointing offset, but were typically between 3 and 10, with higher SNR toward the core's dust continuum peak.    

\subsection{Infall/Expansion Models}
Idealized radiative transfer models that reproduce the spectral asymmetries characteristic of contracting cores have been created so that the infall or expansion velocities of a given core can be extracted from its observed spectra.  The two most widely used spectral-line models are the ``two-layer'' model from \cite{1996ApJ...465L.133M} and the more recent ``HILL5'' model from \cite{2005ApJ...620..800D}. Although these models are similar, there are slight differences between the two arising from assumptions made about core structure.  Both assume there are two regions within a core, but they differ in how the excitation temperature increases between those two layers as a function of opacity.  The two-layer model assumes the excitation temperature increases as a step function at the boundary between the two regions, while the HILL5 model assumes the excitation temperature increases linearly up to a peak at the boundary and then decreases linearly back down to the initial temperature.  The equations that represent each model are both composed of five free parameters. For the two-layer model, these parameters are: (1) the rear excitation temperature (T$_r$) (i.e., excitation temperature of the layer farthest from our point of view), (2) the velocity dispersion of the molecular tracer ($\sigma$), (3) the optical depth of the molecular tracer ($\tau$) (i.e., the opacity at which the emission originates), (4) the velocity of the cloud with respect to the Local Standard of Rest (v$_{LSR}$), and (5) the infall velocity of the system (v$_{in}$).  For the HILL5 model, the only difference is that T$_r$ is replaced by the \emph{peak} excitation temperature (T$_{peak}$).

We chose to use the HILL5 model to obtain infall or expansion velocity estimates because the excitation profile it adopts is likely a better representation of the observed excitation conditions in dense cores. Fitting the HILL5 model to all our spectra, however, implicitly assumes the core is undergoing infall or expansion at all observed positions, with no multiple velocity components along the line of sight.  The first assumption appears to be reasonable considering previous studies have found all three cores in our survey to exhibit extended infall or expansion motions \citep[e.g.,][]{2001ApJS..136..703L,2006ApJ...636..952W, 0004-637X-774-1-20}.  Additionally, Figure 6 shows observations of the optically thin tracers N$_2$D$^+$(2-1) and N$_2$D$^+$(3-2) towards the dust continuum peaks of each core (N$_2$D$^+$(3-2) was not detected toward L492, so we show instead N$_2$D$^+$(2-1)).  F-test comparisons between the best fitting HILL5 model and a Gaussian model for these spectra prefer the simpler Gaussian in all cases (see \S 3.3 for a more detailed discussion of our F-test procedure).  Our fitting procedure simultaneously fits all hyperfine components to provide an estimate of the total optical depth $\tau_{total}$ (i.e., the sum of the optical depths of each individual hyperfine component).  This means that while $\tau_{total} = 3.4$ for L694-2 in N$_2$D$^+$(3-2), the main hyperfine component (F$_1$ F = 4 5 $\rightarrow$ 3 4), which has a fraction of the total line strength of 17.5$\%$, is indeed optically thin (along with all other hyperfine components which have lower contributions to the total line strength).  Although it is not possible to clearly detect multiple line-of-sight velocity components with N$_2$D$^+$(2-1) and N$_2$D$^+$(3-2) since their line profiles are blends of individual hyperfine components, the fact that these lines are better represented by a Gaussian suggests we are likely viewing single velocity components.  In addition, previous observations of optically thin tracers without hyperfine structure such as C$^{17}$O(1-0) and C$^{18}$O(1-0) by \cite{Crapsi_2004} toward L1521F as well as C$^{34}$S(2-1) and HC$_3$N(9-8) by \cite{Hirota_2006} toward L492 also revealed Gaussian line profiles that support a single line-of-sight velocity component.

\subsection{MPFIT Line Fitting}

Each model was fitted to the spectra using the MPFIT suite of non-linear least squares curve fitting functions \citep{2009ASPC..411..251M}.  MPFIT works by performing a series of iterations in which it slightly adjusts the free parameters of the given model until the best fit to a particular spectrum is obtained.  The fitting process begins using user-defined starting parameters based on the number of free parameters for the input model, which for the HILL5 model are T$_{peak}$, $\sigma$, $\tau$, v$_{LSR}$, and v$_{in}$.  Estimates for these values were obtained from previous infall speed surveys, such as those conducted by \cite{schnee_2013} and \cite{2005ApJ...619..379C}, which found velocity dispersions, peak intensities, local standard of rest velocities, and infall speeds towards the emission peaks of the three cores in this analysis.  MPFIT inserts these initial parameters into the equations of the input model and compares the theoretically produced line profile with that of the actual data.  It then repeats this process, shifting one of the five parameters each time, to minimize the sum of the squares of the errors between the two lines (weighted by the RMS noise measurement for the observed spectrum).  Finally, infall or expansion velocity estimates, along with estimates for the four remaining model parameters, are extracted from the best fits to each spectrum.  1-$\sigma$ statistical errors on these estimates are found by taking the square root of the diagonal terms in the covariance matrix, a standard method for determining the variance of a random variable in statistics.  Since each number in the covariance matrix shows how each model parameter depends upon the other parameters, the square root of the diagonal terms gives the variance of each parameter about its own expectation value (i.e., the spread or dispersion around the mean parameter value over all iterations completed by MPFIT).  Following this methodology, we obtained infall or expansion velocities for all spectra in this survey above an SNR of 6.  The infall or expansion velocity estimates can be seen in Tables 3-5 for each molecular tracer and observed position.  The MPFIT calculated error estimates for the measured infall or expansion velocities can be seen as $\pm$ values, respectively, alongside the velocities of Tables 3-5.  The SNR of each spectrum is also listed in Tables 3-5.  Figures 2-4 display all the spectra obtained in this survey along with their corresponding best fit HILL5 model if a given spectrum is above the SNR threshold (Figures 2b-2d, 3, and 4 are available in the online version of the Journal).

N$_2$H$^+$(1-0) emission is split into seven hyperfine components that can individually exhibit asymmetries characteristic of infalling or expanding cores.  Previous N$_2$H$^+$(1-0) observations of L1544 by \cite{1999ApJ...513L..61W}, \cite{Caselli_2002_1}, and \cite{Bizzocchi_2013} and of L694-2 by \cite{2006ApJ...636..952W} showed evidence for asymmetries in all seven hyperfine structures.  Our N$_2$H$^+$(1-0) spectra also show similar asymmetries in each hyperfine component for all three cores (see, e.g., Figure 5).  These hyperfine components have varying levels of sensitivity to core infall/expansion motions, however, due to the different excitation conditions traced by each line.  For instance, the central F$_1$ F = 2 3 $\rightarrow$ 1 2 component is the brightest and contains the highest optical depth, while the isolated F$_1$ F = 1 0 $\rightarrow$ 1 1 component is weakest and contains the lowest optical depth.  Hence, we fit only the central hyperfine component of N$_2$H$^+$(1-0) to derive the infall/expansion velocities used for our analysis.   



\subsection{F-test Comparison of HILL5 and Gaussian Models}
We also fit a Gaussian model to each spectrum and compare these fits to those of the HILL5 model with F-tests.  The F-test takes into account the $\chi^2$ values and degrees of freedom for each fit to determine if a simpler or more complex model better represents a set of data.  We chose the null hypothesis that the HILL5 model does not provide a better fit than the Gaussian model.  Table 6 shows the percentage of spectra with SNR$>$6 in each core and transition that reject the null hypothesis at the 90$\%$ confidence level (i.e., the HILL5 model is preferred over the Gaussian model).  Overall, we see that the HILL5 is preferred by the F-test for the majority of spectra included in our analysis.

\subsection{MCMC Line Fitting}
Although MPFIT provides estimates for the best fit HILL5 model parameters based on the non-linear least squares approach, it fails to produce similarly likely solutions that may exist at slightly larger $\chi^2$ values.  Least squares minimization routines are susceptible to falling into global minima (or in some cases local minima) solutions without acknowledging the existence of additional minima solutions.  To explore the existence of these solutions, we conducted a line fitting procedure utilizing the Metropolis-Hastings Markov Chain Monte Carlo method, which periodically forces itself out of global and local minima solutions to produce a probability distribution over multiple solution sets.  We chose the traditional weighted least squares equation as our error function and a Gaussian likelihood function, forcing the likelihood to be maximal when the error function is minimal.  Our Metropolis-Hastings algorithm begins by calculating the value of the likelihood function for an initial guess of the parameters in the model.  It then proposes a new set of parameters by randomly drawing values between $\pm$0.005 that are added to the initial guess parameters.  Next, it compares the ratio of the newly proposed parameter set's likelihood function value to that of the initial parameter set's likelihood function value.  If this likelihood ratio is larger than a randomly drawn number between 0 and 1, the initial parameter set is replaced by the proposed parameter set and recorded as a minimal likelihood set of parameters.  Conversely, if the likelihood ratio is less than the randomly drawn number between 0 and 1, the proposed set of parameters is not recorded and the process is repeated with an altered proposed parameter set, as discussed above.  The algorithm is repeated iteratively until 10$^6$ minimal likelihood parameter sets have been recorded.  For consistency, our initial guess parameter set is the same as the initial guess used for the MPFIT line fitting procedure.  

Figure 7 displays the probability mass functions of v$_{in}$ obtained by completing the aforementioned MCMC line fitting procedure for four DCO$^+$(2-1) spectra, with varying levels of SNR, observed in L694-2.  Overplotted in Figure 7 are the MPFIT best fit v$_{in}$ estimates along with their corresponding uncertainty.  We find that despite the existence of local minima likelihood solutions for the low SNR cases, the best fit value of v$_{in}$ found by MPFIT is consistent with the most probable value of v$_{in}$ determined by the MCMC method in all four cases.  Moreover, the MPFIT uncertainty estimates are similar to those that would be obtained by taking the standard deviation of each distribution's most probable v$_{in}$ peak (disregarding the less likely solution peaks).  For these reasons, the analysis presented in the following sections uses the MPFIT v$_{in}$ estimates along with their corresponding 1-$\sigma$ statistical uncertainties calculated from the MPFIT parameter covariance matrix.  We also note that the less likely solution peaks occur at lower values of v$_{in}$ in both low SNR cases.  If these lower probability peaks were used for our analysis, they would likely increase any positional variations observed in v$_{in}$ since the low SNR spectra are generally pointings with larger radial offsets from the core dust continuum peak.



\section{Results and Discussion}

Using the HILL5 infall or expansion estimates, Figure 8 shows the 12 velocity maps constructed for all cores and lines observed.  Positive velocities (yellow, orange, and red) correspond to infall, while negative velocities (green and blue) represent expansion.  Since spectra with SNR $<$ 6 were omitted to exclude velocity estimates with high uncertainties, there are differences in the mapped coverage across the four transitions.  DCO$^+$(3-2) and HCO$^+$(3-2) had the noisiest spectra in our sample, which caused their final coverage to be diminished compared to that of N$_2$H$^+$(1-0) and DCO$^+$(2-1).  

In Figure 9, we show infall velocity versus radial offset from the map dust continuum peak for each line with SNR $>$ 6 and v$_{in}$ uncertainties less than 0.3 km s$^{-1}$ for all three cores.  The radial offsets in Figure 9 are calculated using the map's dust continuum peak as the center of the core due to the $\Delta\alpha$=0, $\Delta\delta$=0 pointings in L492 and L1521F being offset from the (likely) core center (see Figure 8, which plots the dust continuum peak position as an ``x" and the $\Delta\alpha$=0, $\Delta\delta$=0 pointing as a ``+" in each map). 

We investigate infall as a function of molecular tracer by comparing the median infall velocities for all pointings observed in a single transition.  Table 7 shows median v$_{in}$ estimates for each transition using all spectra with SNR $>$ 6.  We discuss in the following subsections the observed trends in these values.


\subsection{Infall Positional Variations}
We observe variations of infall velocity with position across all three cores.  As seen in Figure 9, which plots v$_{in}$ versus core radius for a given pointing, positional variations are more apparent for N$_2$H$^+$(1-0) and DCO$^+$(2-1) observations when compared to the other molecules.  The DCO$^+$(2-1) pointings near the dust continuum peaks in L492, L694-2, and L1521F produce faster values of v$_{in}$, by approximately 50 m~s$^{-1}$, when compared to the pointings at larger radii.  N$_2$H$^+$(1-0) also produces faster infall velocities toward the center of L1521F, with a difference of $\sim$40 m~s$^{-1}$ from the slower outer pointings.  These positional variations are significant considering the individual uncertainties for each velocity measurement shown in Figure 9 are generally less than 10 m~s$^{-1}$.  Unlike \cite{2006ApJ...636..952W}, we do not see a clear trend of decreasing speeds at larger core radii for L694-2 in N$_2$H$^+$(1-0).  This difference could be due to the fact that \cite{2006ApJ...636..952W} fit the HILL5 model to averaged spectra over annuli at varying core radii with a high spatial resolution of 10''.  Our lower spatial resolution of 27'' leads to overlapping pointings in L694-2, which may increase the difficulty of detecting positional variations in v$_{in}$ when combined with our single-pointing fitting method.  The higher spatial resolution of our DCO$^+$(2-1) observations reduces the amount of overlap between pointings, which is likely the reason we see a larger difference between the central and outer pointings for L694-2 in this transition.  While DCO$^+$(3-2) and HCO$^+$(3-2) show positional variations over 100 m~s$^{-1}$ for L694-2, the presence of lower infall speeds at larger core radii is less clear for these transitions.

As we intend here to investigate the biases involved in single-pointing, single-tracer infall surveys, creating 3D models of these cores and testing our results against those predicted by core contraction theories is outside the scope of this paper.  Our results clearly show that as long as a single-pointing optically thick line observation is made near the dust continuum peak of a prestellar core, it may indicate the core as either infalling or expanding.  This claim is bolstered by our observations of both L492 and L694-2, which show inward motions in nearly all central pointings for all selected transitions.  The same approach cannot be done for protostellar cores since bipolar outflows can cause expansion near the core center, as observed in L1521F.  Despite the outflow in L1521F being moderate compared to others seen in various low- to high-mass star-forming regions, we still detect it in our data.  This result suggests that outflows can cause detectable expansion motions even in the earliest protostellar stage ($<$10$^4$ yr).




\subsection{Infall/Expansion Molecular Variations}

Given the four tracers used in the survey, one may expect based on the inside-out collapse model that N$_2$H$^+$(1-0) would produce the fastest inward motions since it traces the higher density central regions where infall is theorized to be largest.  Similarly, HCO$^+$(3-2), which is thought to be a lower density tracer, would be expected to yield the slowest speeds since it represents the sparser outer layers where CO can survive in the gas phase and where infall is expected to be slowest. DCO$^+$ is somewhat of an oddball because it contains both deuterium, which traces central regions, and also CO, which traces outer regions.  Therefore, DCO$^+$ could trace a middle ground within cores where both deuterium and CO are found in the gas phase.  A discrepancy in the measurements of v$_{in}$ may also be expected between the DCO$^+$(2-1) and DCO$^+$(3-2) transitions because the latter has a higher critical density than the former.  In addition, more pronounced absorption dips may be expected in the spectra of DCO$^+$(2-1) versus DCO$^+$(3-2), since there is a higher probability that the emission of the former will be reabsorbed.  

Our results show that choice of molecular tracer plays an important role for infall measurements, as the ranges of v$_{in}$ magnitudes vary amongst the four observed molecular lines (see, e.g., Figures 8 and 9).  Table 7 shows that median infall speeds vary up to 65 m~s$^{-1}$ between certain transitions.  These variations appear significant when taking into account their respective $\sigma$/$\sqrt{N}$ uncertainties, which are on the order of 10 m~s$^{-1}$ for all transitions.  Table 7 also shows that the tracers producing the fastest or slowest infall speeds are highly dependent on the particular core being observed.  No single tracer consistently produced the fastest (or slowest) speeds for all cores.  This result indicates that a core's dynamical and chemical evolutionary state may impact an observer's ability to trace accurately infall or expansion motions.

Of the three cores observed, L1521F appears to match the chemical dynamics of cores expected from inside-out collapse, with only DCO$^+$(2-1) being out of place by having the largest infall velocities.  It may be that DCO$^+$ is actually found deeper within this core than we first surmised as a result of its advanced evolutionary stage. Indeed, due to heating caused by the embedded forming protostar in L1521F, CO could be returning to the gas phase at higher densities and producing DCO$^+$ within deeper regions of the core, resulting in the faster-than-expected infall velocities observed from its transitions.  Although the discrepancy between the two transitions appears insignificant considering their uncertainties, if real, it possibly arises from the outer core layers preferentially reabsorbing the DCO$^+$(2-1) emission.  Heating from the embedded protostar likely produces DCO$^+$(2-1) and DCO$^+$(3-2) emission over similar regions within the high density central parts of the core, but the outer lower density layers of the core may not have as much DCO$^+$ excited to the J=2 state than the J=1 state.  As the DCO$^+$(2-1) and DCO$^+$(3-2) emission pass through the outer core layers in-between our line of sight and the central regions from which they originate, the higher amount of DCO$^+$ molecules in the J=1 state would cause relatively more reabsorption of the DCO$^+$(2-1) emission, leading to larger asymmetries in its line profiles that produce higher infall speed estimates.  Radiative transfer modelling of L1521F may provide more insight into whether or not the outer layers are the cause of the higher infall velocities for DCO$^+$(2-1), but with only two DCO$^+$ transitions observed in this study, and only four transitions observed in total, our limited dataset would likely be insufficient for such a project.     

In L694-2, we see significantly different behavior.  HCO$^+$ produces the largest  speeds while N$_2$H$^+$ yields the lowest and DCO$^+$ lies in the middle of the range.  One explanation could be that L694-2 is still accreting material from the surrounding molecular cloud, but it has not yet built large enough mass to start contracting as quickly in the denser regions.  This scenario explains why we see the fastest infall for the presumed outer layer tracer HCO$^+$ and slowest speeds for the presumed inner layer tracer N$_2$H$^+$.  These results may also indicate that L694-2 is less chemically evolved than L1521F.  Since the region of the core where particular molecules are most abundant is relative to core evolutionary stage, the tracers used in this survey may not be representing exactly the same areas in each of the cores.  From previous studies, we know that L1521F is protostellar and thus farther along its dynamical evolution than L694-2.  Our observations match this fact since L1521F velocities show some correspondence with behavior predicted based upon the current understanding of core chemistry.

In L492, the weaker DCO$^+$(3-2) and HCO$^+$(3-2) lines result in only single v$_{in}$ estimates that meet our SNR threshold for these tracers.  Both of these spectra, however, have noisy redshifted components that were not fit during our fitting procedure but may indeed be real emission line features.  If these noisy redshifted components are included in our fitting procedure, the resulting DCO$^+$(3-2) v$_{in}$ estimate increases to 0.111 $\pm$ 0.016 km s$^{-1}$ and the resulting HCO$^+$(3-2) v$_{in}$ estimate increases to 0.297 $\pm$ 0.013 km s$^{-1}$.  Thus, it is more difficult to compare these tracers to N$_2$H$^+$(1-0) and DCO$^+$(2-1) for this core, which means the forthcoming discussion concerning the chemical evolution of L492 is highly speculative.  Higher SNR DCO$^+$(3-2) and HCO$^+$(3-2) spectra across multiple pointings in L492 need to be obtained in the future to determine if our speculation is an accurate analysis of this core.  Nevertheless, the currently available data for L492 reveal that DCO$^+$ produces faster infall speeds than N$_2$H$^+$.  If our individual HCO$^+$(3-2) v$_{in}$ estimate is indeed indicative of slower infall across the core in that tracer, it would suggest L492 has chemical behavior comparable to L1521F. These results may point toward L492 being in an intermediary evolutionary stage in-between L1521F and L694-2, in which it is more chemically evolved than L694-2 but has yet to ignite a protostar at its center.  

We can use other measures, such as deuterium fractionation, to provide further insight into the relative evolutionary state of each core.  Analyses by \cite{2005ApJ...619..379C} found that L694-2 had much higher deuterium fractionation than L492, which is thought to be an indicator of dynamical evolution in starless cores, with the former having [N(N$_2$D$^+$)/N(N$_2$H$^+$)] = ${0.26~{\pm}~0.05}$ and the latter having [N(N$_2$D$^+$)/N(N$_2$H$^+$)] = ${0.05~{\pm}~0.01}$.  As an additional measure of dynamical evolution, we also calculated peak H$_2$ column densities for each core using their peak dust continuum fluxes from Figure 1.  We assumed a uniform dust temperature of 10 K for all cores, dust opacity law of $\kappa _\nu$ = 0.1$\times$($\lambda$/300$\mu$m)$^{-\beta}$ cm$^{-2}$/g, and fixed the dust emissivity index $\beta$ to 2.  We find the H$_2$ column densities (in cm$^{-2}$) for L1521F, L694-2, and L492 to be 1.4$\times$10$^{23}$, 9.0$\times$10$^{22}$, and 1.9$\times$10$^{22}$, respectively.  Since column density is thought to increase as cores evolve, this metric also indicates that L492 is less evolved than L694-2.  

While all three cores show signatures of infall and are thus dynamically evolved in that measure, their chemical evolutionary stages may be quite different.  \cite{Hirota_2006} show that individual cores can exhibit both chemically evolved and young signatures depending on the type of evolutionary tracer observed (e.g., CO depletion, infall kinematics, and gas phase molecular abundance fractions).  L492 appears to be a core that falls into this category, displaying both young and evolved chemical signatures.  If L1521F, L694-2, and L492 are indeed in different stages of the star formation process, our results may show that the core layers traced by a certain molecule/transition vary as the core evolves.  Thus, when conducting a single-tracer or single-transition survey of core infall, one must consider the individual evolutionary state of each observed core to understand the specific core layers being traced in each source.  The environment surrounding a core undoubtedly also plays a role in gas phase abundance measurements.  For instance, L492 lies in the Aquila Rift region where external irradiation has caused on average higher dust temperatures than more isolated regions \citep[see, e.g.,][]{Konyves_2015}.  The warmer medium surrounding L492 could explain its lower deuterium fraction compared to the other cores since warmer temperatures would help prevent CO freeze-out, lowering the production of deuterated molecules.  

Detailed radiative transfer chemical modelling of cores is used to gain insight into the abundances of particular molecules and the chemical processes occurring inside cores, such as molecular differentiation \citep{Tafalla_2002, Bergin_2002} and core age \citep{Maret_2013}.  While this additional information may help unravel the causes of the molecular dependence we observe, additional modelling is outside the scope of this paper.  Our results suggest, regardless of the physical and chemical mechanisms responsible for the observed variations in contraction speeds, that measured infall or expansion velocities indeed depend on the observed transition.  Median velocity measurements from a core vary by $\sim$ 30 m~s$^{-1}$ or more between some transitions, which is significant considering their respective uncertainties are $\sim$ 10 m~s$^{-1}$, indicating that single-tracer observations of infall or expansion are dependent on their choice of molecular tracer.  To understand fully the \textit{rate} at which a core is either infalling or expanding, one must observe multiple transitions from multiple molecules.  If one's goal is to simply determine if a dense core is either infalling or expanding, regardless of the actual speed, then a single transition is sufficient.  This statement may also hold for detecting bipolar outflows in a protostellar core because the expansion motions we observe in L1521F appear to be located in the same regions of the core in at least two of the observed transitions (DCO$^+$(2-1) and N$_2$H$^+$(1-0)).


\subsection{Expansion Motions}
Several examples of expansion motions were also detected in this analysis, along the outer regions of all three cores.  Figure 5 shows examples of asymmetrically red spectra, indicating expansion, observed in all three cores.  Figure 5 also provides a comparison between spectra that have been identified as asymmetrically red versus blue in each core.  Expansion motions were to be expected in L1521F, considering it has been classified as protostellar and a bipolar outflow has been detected in previous observations \citep{0004-637X-774-1-20}.  Our measurements appear to confirm this bipolar outflow, as can be seen in the right column of Figure 8, which indicates expansion motions along the south-east and north-west ends of the core.  On the other hand, the expansion motions observed in L492 and L694-2 are surprising since these cores have been classified as prestellar.  Both N$_2$H$^+$(1-0) and DCO$^+$(3-2) indicate expansion motions along the outskirts of L694-2 (see middle column of Figure 8) within the range of -0.067 km~s$^{-1}$ to -0.010 km~s$^{-1}$, but these estimates contain high uncertainties due to the decreased SNR of the spectra from which they were obtained.  Similarly, both N$_2$H$^+$(1-0) and DCO$^+$(2-1) show expansion motions within the range of -0.098 km~s$^{-1}$ to -0.004 km~s$^{-1}$ for several outer pointings in L492 (see left column of Figure 8).  Similar outward motions have been observed in prestellar cores by previous surveys such as \cite{2007ApJ...664..928S}, \cite{2011ApJ...734...60L}, and \cite{schnee_2013}.  Since L492 and L694-2 are classified as starless, it is doubtful that this expansion is related to an outflow jet.  Instead, these cores may be in an oscillatory state in which the low density material in the outer layers is expanding while the majority of the core continues to contract and show inward motions.  Oscillatory cores of this nature have been observed in previous surveys \citep[e.g., ][]{0004-637X-586-1-286}, and may be caused by perturbations induced by turbulence originating from the parental molecular cloud or nearby supernovae feedback.




\section{Future Work}
Although only three cores and four transitions were analyzed in this paper, our knowledge of the early stages of the star formation process could be improved if infall/expansion maps were created for more cores using a wider variety of transitions.  Expanding our survey to include additional prestellar sources, such as L1197 and Oph D, as well as protostellar sources, such as L429 and L328, would allow us to observe the effects that environment has upon collapse kinematics and chemistry.  These data would also allow us to determine whether the expansion motions found on the outskirts of L492 and L694-2 are common in other cores or if they are a characteristic that is rarely observed.  Additionally, the velocity patterns of L1521F could be compared to those of other protostellar sources to characterize the bipolar outflow process and its impact on calculated infall or expansion motions.  Observing additional molecular line transitions that trace a wider range of excitation conditions, such as the high density tracers N$_2$D$^+$(3-2) and N$_2$H$^+$(3-2), as well as the low density tracers HCN and CS, would provide a better overall picture of prestellar and protostellar infall as a function of core depth.  Using the critical densities of each molecular tracer, data cubes which include infall/expansion as a function of both core position and depth could then be created.  These cubes would provide a three-dimensional interpretation of core contraction, from which an improved infall/expansion model could be developed.  For instance, current prestellar core collapse models, such as the HILL5 and two-layer, fail to take into account either the molecular tracer used for the observations or the position on the core where measurements were obtained.  A radiative transfer model that considers these additional parameters may provide more accurate spectral line fitting, allowing for more precise infall or expansion velocity measurements in future studies.

\section{Summary}
Theory predicts that the speed at which starless cores contract is dependent upon the distance from the core center, while chemical dynamics suggest that contraction is also dependent upon the observed molecular tracer.  Yet few surveys of core infall have combined multi-pointing, multi-transition observations to measure infall velocities.  To investigate the biases involved in single-pointing/single-transition infall surveys, we observed multiple positions in three cores (two starless and one protostellar) using four molecular tracers with similar values of n$_{cr}$ (N$_2$H$^+$(1-0), DCO$^+$(2-1), DCO$^+$(3-2), and HCO$^+$(3-2)) and determined infall or expansion velocities based on the asymmetries of the obtained spectra.  We find that infall velocities do indeed vary with both the observed position and molecular tracer.  Estimated line-of-sight infall speeds traced by DCO$^+$(2-1) decrease by 40-50 m s$^{-1}$ when observing at least $\sim$0.04 pc from the dust continuum emission peak for all three cores.  In L1521F, infall speeds estimated from N$_2$H$^+$(1-0) spectra also show a decrease of $\sim$40 m s$^{-1}$ for pointings at core radii larger than 0.04 pc.  Median infall speeds calculated from all observed positions across a core with SNR $>$ 6 also vary by as large as 65 m s$^{-1}$ depending on the choice of molecular tracer.  Both prestellar cores (L492 and L694-2) show overall signs of inward motions, with expansion motions being detected only on the outer positions in some transitions.  The protostellar source (L1521F) also showed inward motions, with the exception of bipolar outflows along one axis.     

These results suggest that both position, molecular tracer, and evolutionary state must be taken into consideration when attempting to characterize the overall rate at which a core is contracting.  If the goal of a particular study is to simply determine whether a prestellar core is contracting or stagnant, however, a single-pointing observation using one molecular tracer may be sufficient as long as the observed position is somewhere near the core's dust continuum emission peak.  When observing protostellar sources, this approach does not apply due to the possibility of multiple line-of-sight motions by bipolar outflows.

J.K. acknowledges funding from the National Science Foundation for the portion of this work completed during his participation in the NRAO REU program.  J.K. also acknowledges support from the William Marshall Bullitt Fund, the Henry Vogt scholarship program, and the Bennett Memorial Fund for the portion of this work completed as part of his Honors Thesis at the University of Louisville.  RF is a Dunlap Fellow at the Dunlap Institute for Astronomy \& Astrophysics.  The Dunlap Institute is funded through an endowment established by the David Dunlap family and the University of Toronto.

The National Radio Astronomy Observatory is a facility of the National Science Foundation operated under cooperative agreement by Associated Universities, Inc.  The IRAM 30m telescope is supported by INSU/CNRS (France), MPG (Germany) and IGN (Spain).

{\it Facility:} \facility{IRAM: 30m}

\bibliographystyle{apj}
\bibliography{adssample}

\begin{figure}[ht]
\epsscale{0.5}
\plotone{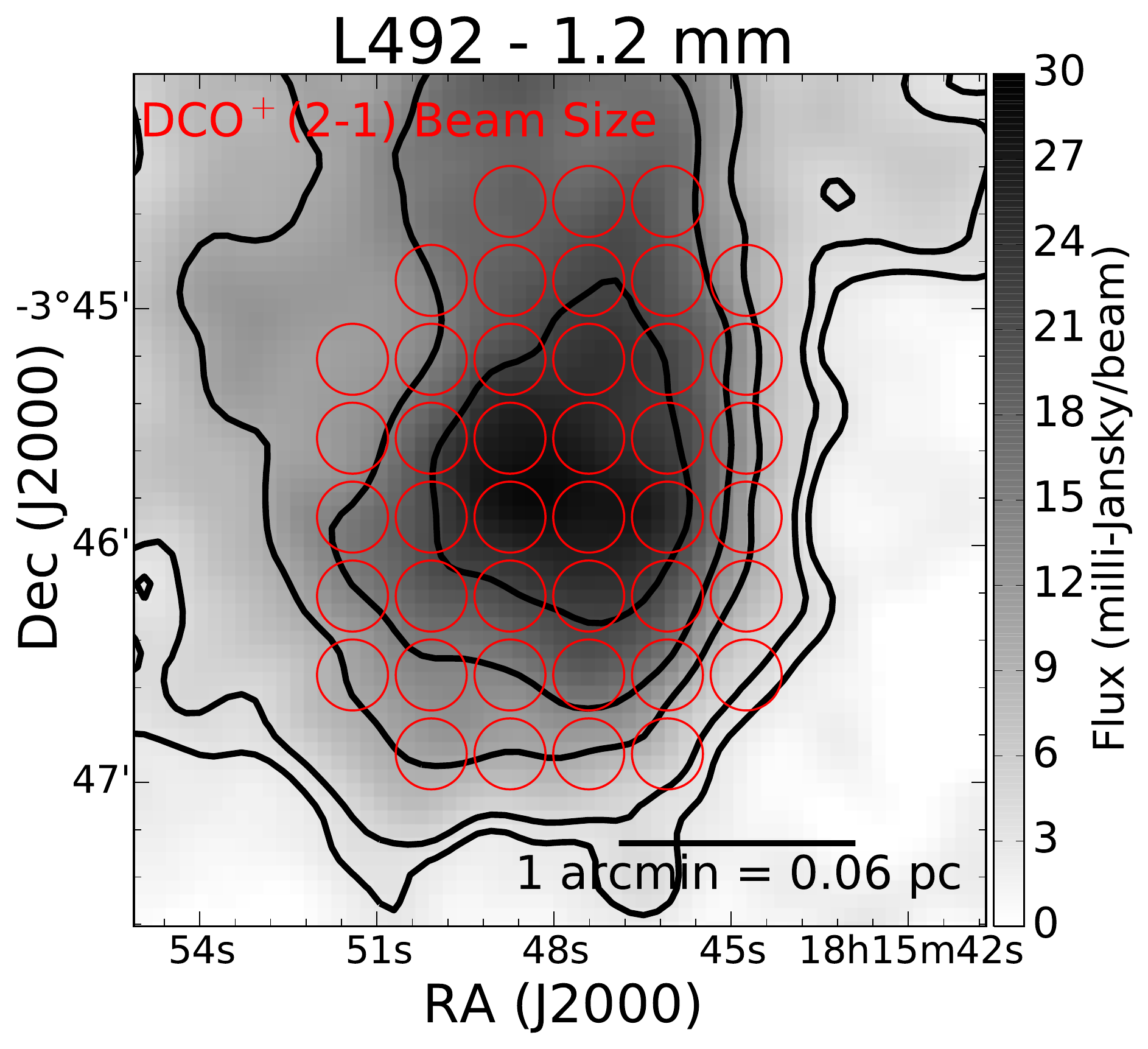}
\centering
\epsscale{1.1}
\plottwo{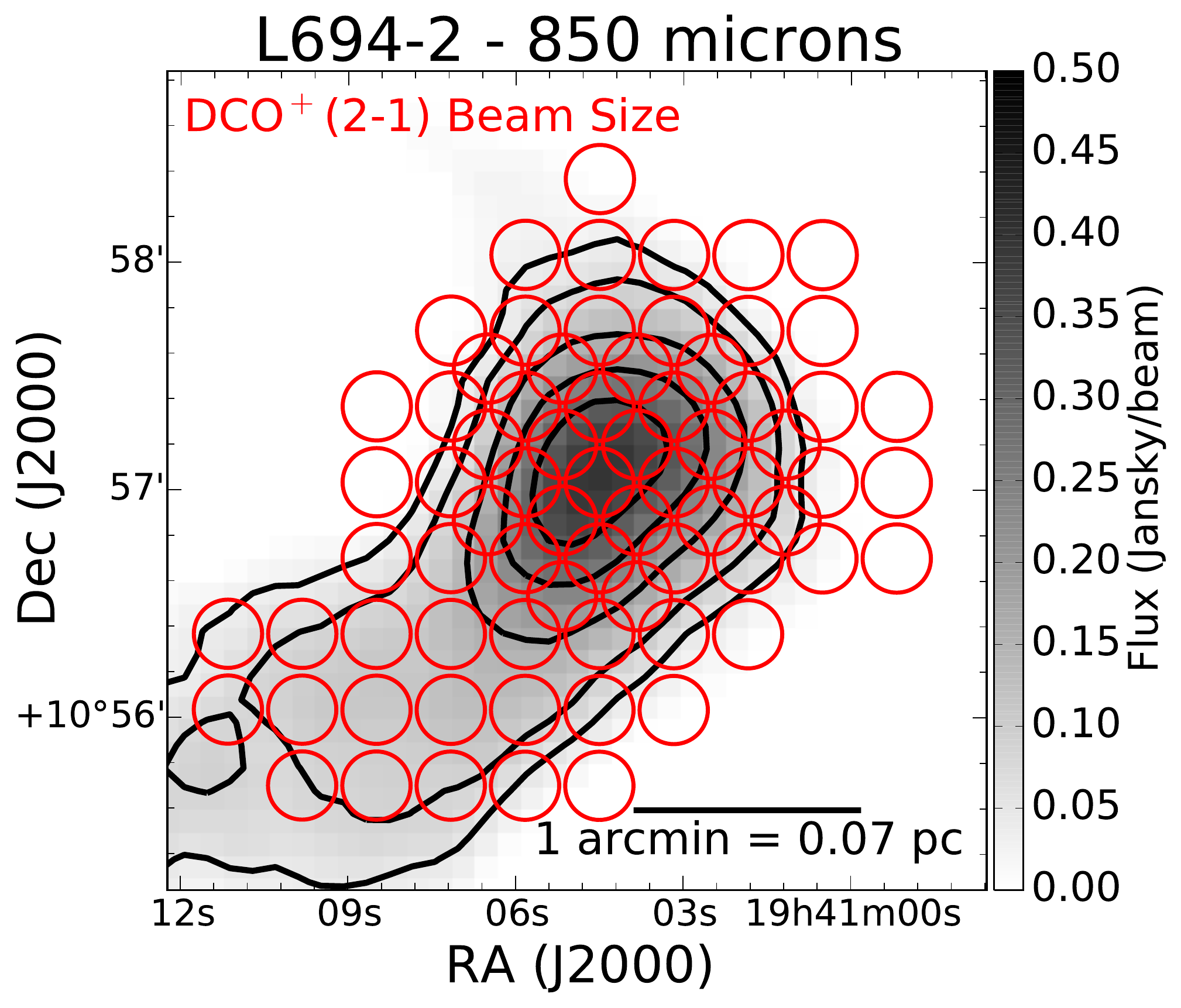}{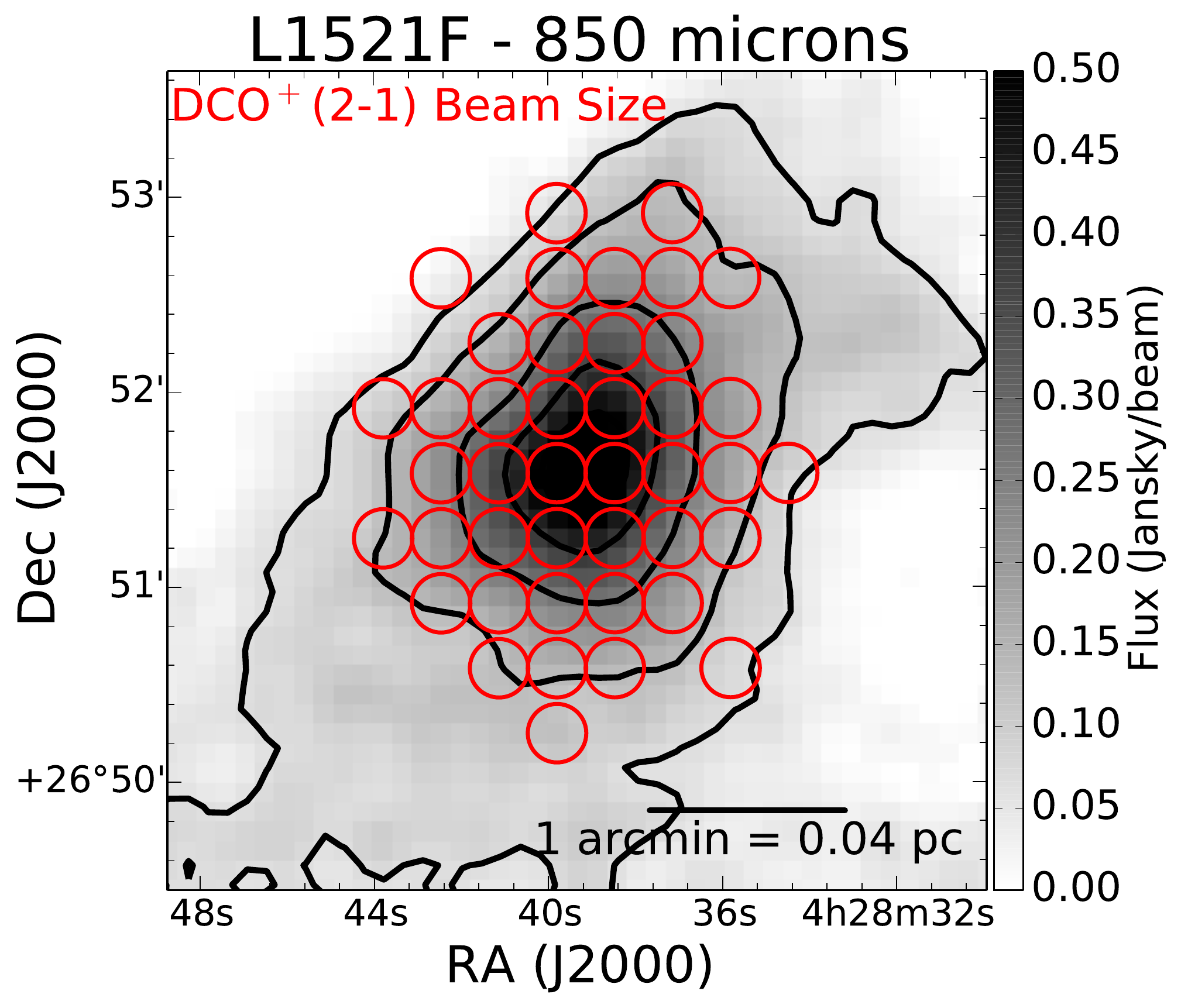}
\caption{Overview of the observations obtained for this survey.  Background is MAMBO 1.2~mm dust continuum emission for L492 \citep[top;][]{kauffmann_2008} and SCUBA 850~$\mu$m dust continuum emission \citep{0067-0049-175-1-277} for L694-2 (bottom left) and L1521F (bottom right) showing roughly the column density structure of each core.  Darker gray corresponds to more emission and therefore likely higher column density, and density.  The contours represent intensities 70, 50, 35, 15, and 10 percent of the peak.  The red circles correspond to the points at which spectra were measured and their size represents the DCO$^+$(2-1) beam size.}
\end{figure}


\begin{sidewaysfigure}[ht]
\figurenum{2a}
\includegraphics[scale=0.66]{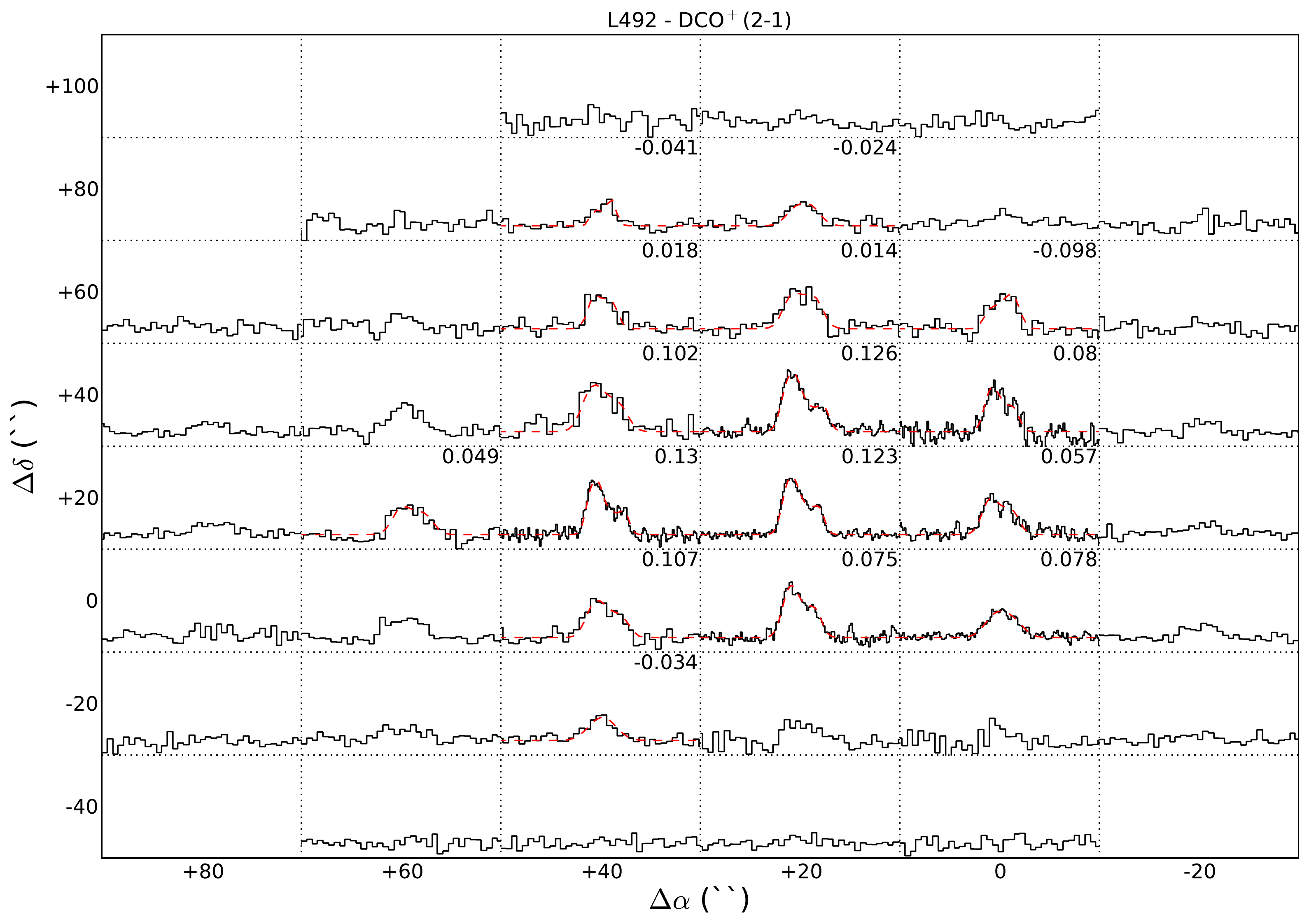}
\caption{Observed DCO$^+$(2-1) spectra (black) for the core L492 at each pointing in this survey.  X and Y axes show the offset in arcseconds from the map's central pointing of 18:15:46.08, -03:46:12.8 (J2000) where a spectrum was obtained.  All spectra are centered on 6.5 $\leq$ V$_{LSR}$ (km s$^{-1}$) $\leq$ 9.0 and -0.5 $\leq$ T$_B$ (K) $\leq$ 3.0.  The red dashed line shows the best fit HILL5 model to the observed spectrum if the latter has SNR $>$ 6.  The number in the upper right corner of each spectrum shows the estimated v$_{in}$ in km s$^{-1}$ based on the best fit model. Similar plots for the other molecular tracers and cores in this survey are available in the online version of the Journal.}
\end{sidewaysfigure}

\begin{sidewaysfigure}[ht]
\figurenum{2b}
\includegraphics[scale=0.66]{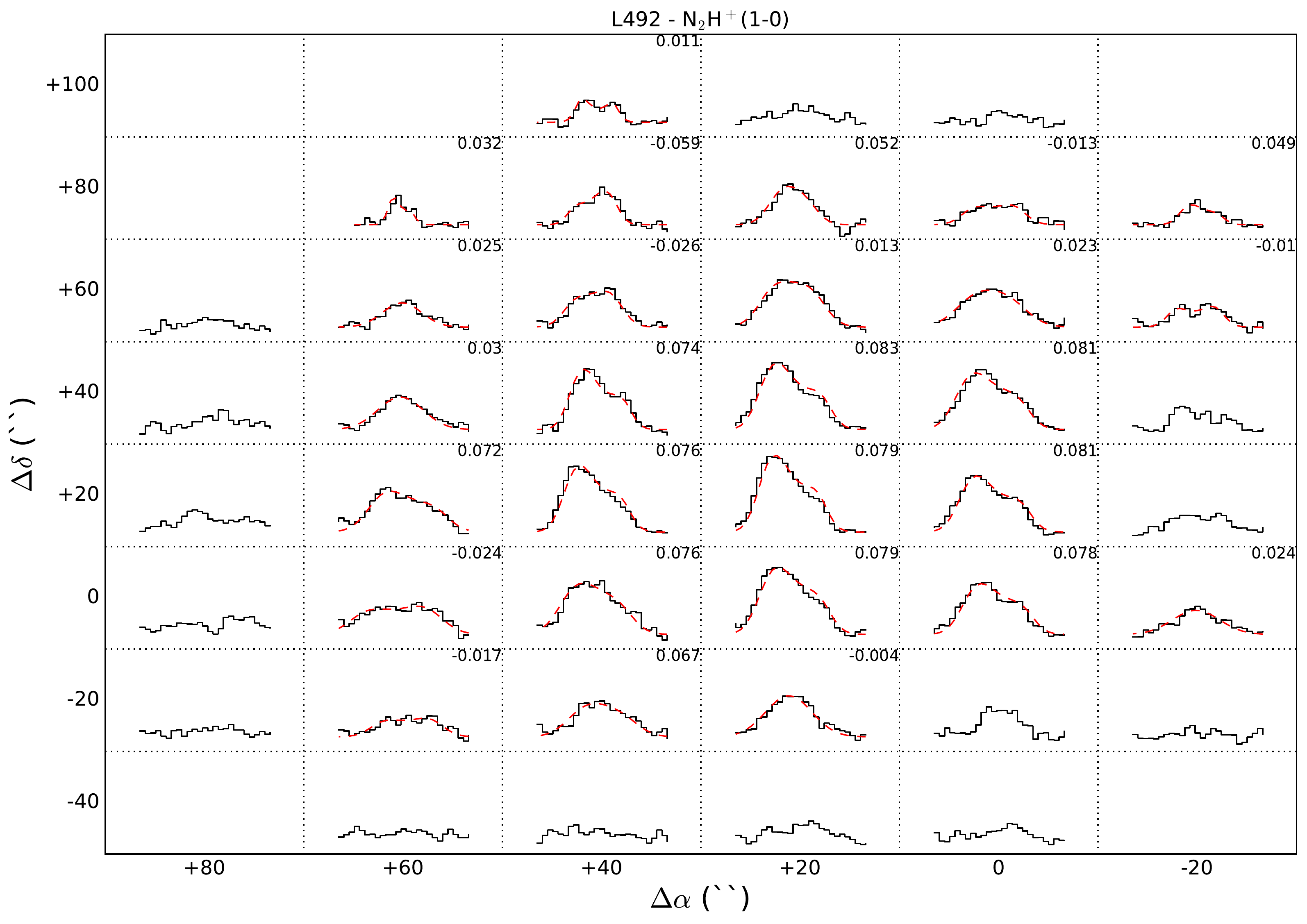}
\caption{Same as Figure 2a using the tracer N$_2$H$^+$(1-0).  All spectra are centered on 7.2 $\leq$ V$_{LSR}$ (km s$^{-1}$) $\leq$ 8.4 and -0.5 $\leq$ T$_B$ (K) $\leq$ 3.0.  Only the central hyperfine structure is displayed in each plot.}
\end{sidewaysfigure}

\begin{sidewaysfigure}[ht]
\figurenum{2c}
\includegraphics[scale=0.66]{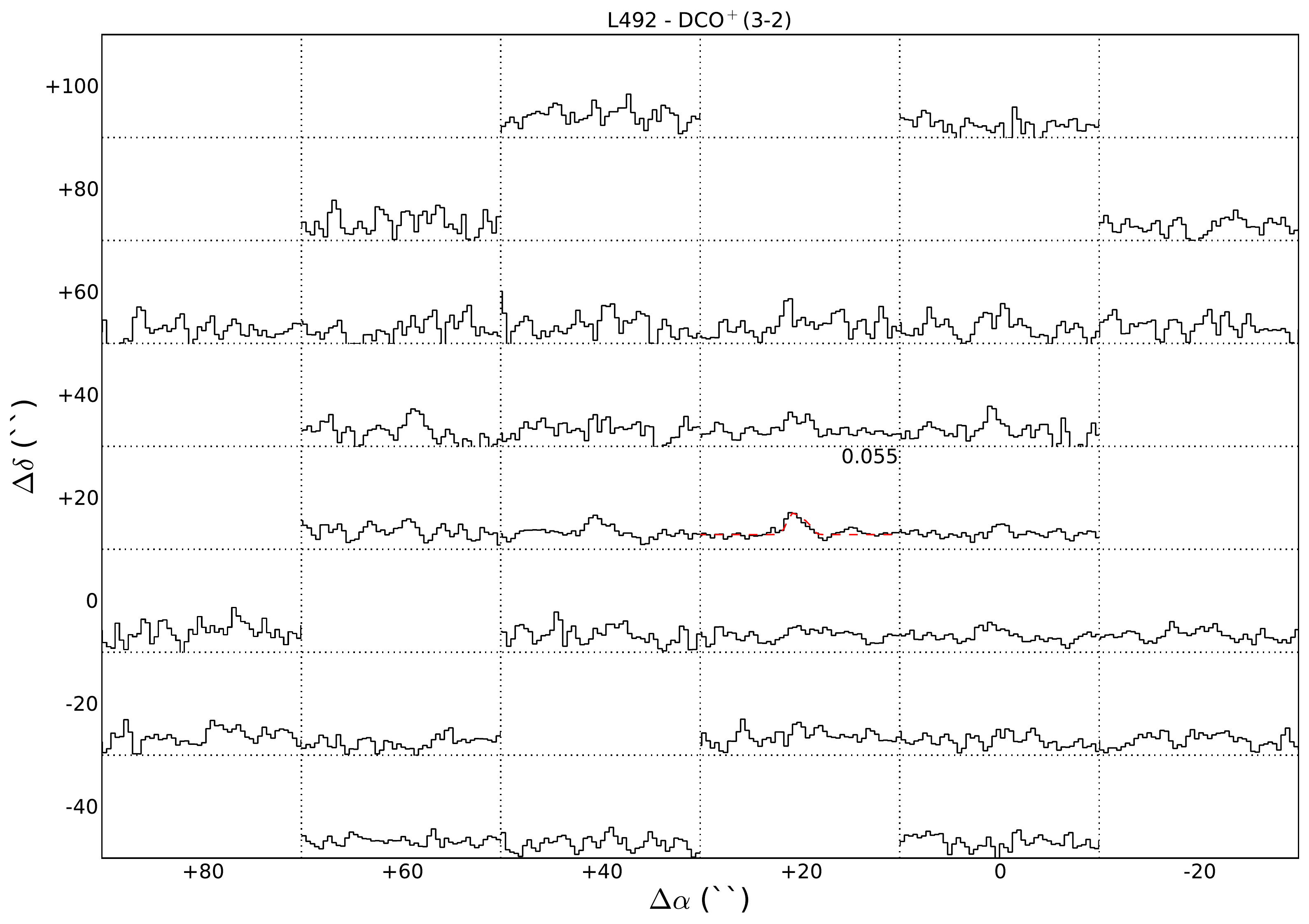}
\caption{Same as Figure 2a for the tracer DCO$^+$(3-2).  All spectra are centered on 6.5 $\leq$ V$_{LSR}$ (km s$^{-1}$) $\leq$ 9.0 and -0.5 $\leq$ T$_B$ (K) $\leq$ 3.0.}
\end{sidewaysfigure}

\begin{sidewaysfigure}[ht]
\figurenum{2d}
\includegraphics[scale=0.66]{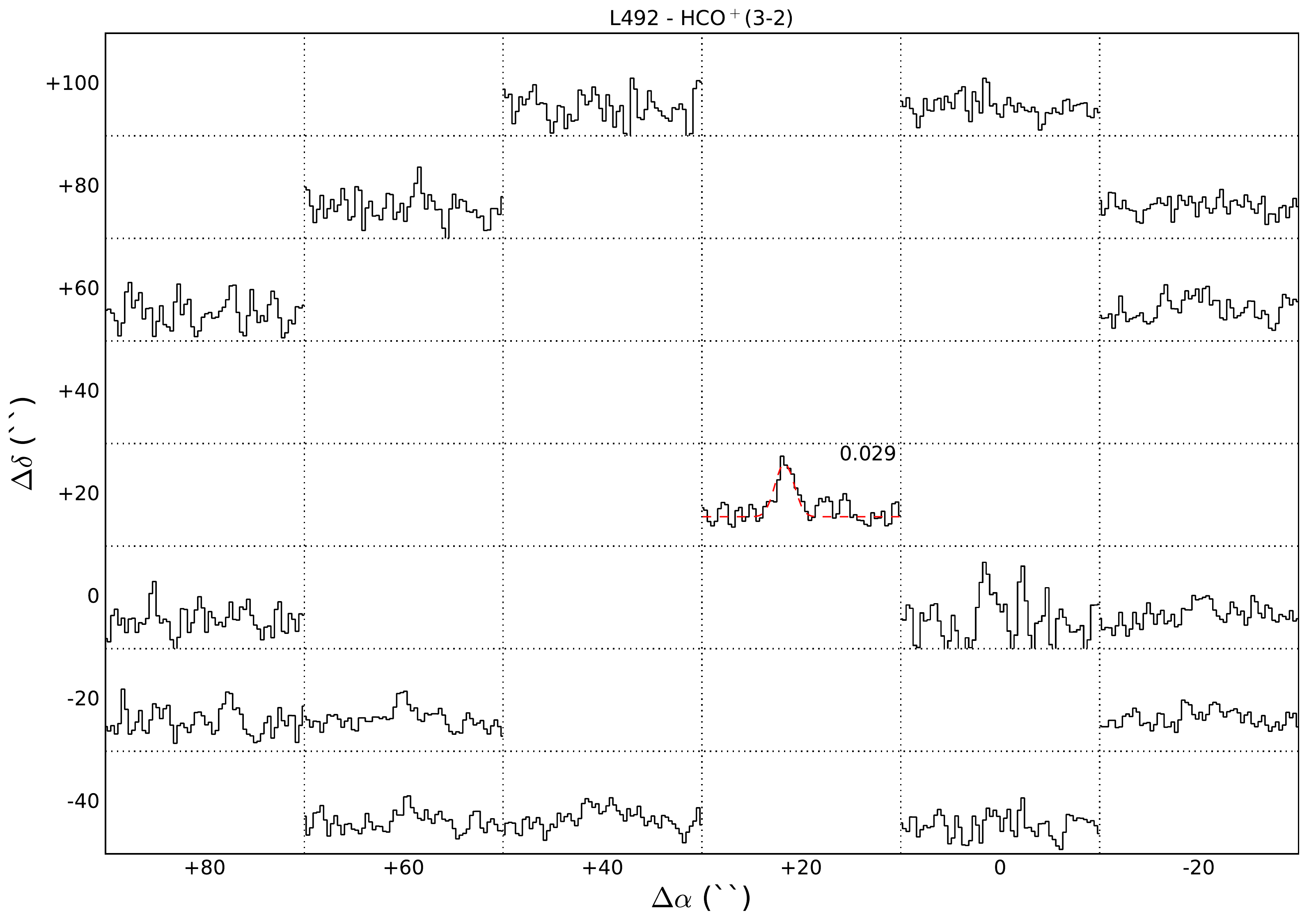}
\caption{Same as Figure 2a for the tracer HCO$^+$(3-2).  All spectra are centered on 6.5 $\leq$ V$_{LSR}$ (km s$^{-1}$) $\leq$ 9.0 and -2.0 $\leq$ T$_B$ (K) $\leq$ 5.0.}
\end{sidewaysfigure}

\begin{sidewaysfigure}[ht]
\figurenum{3a}
\includegraphics[scale=0.66]{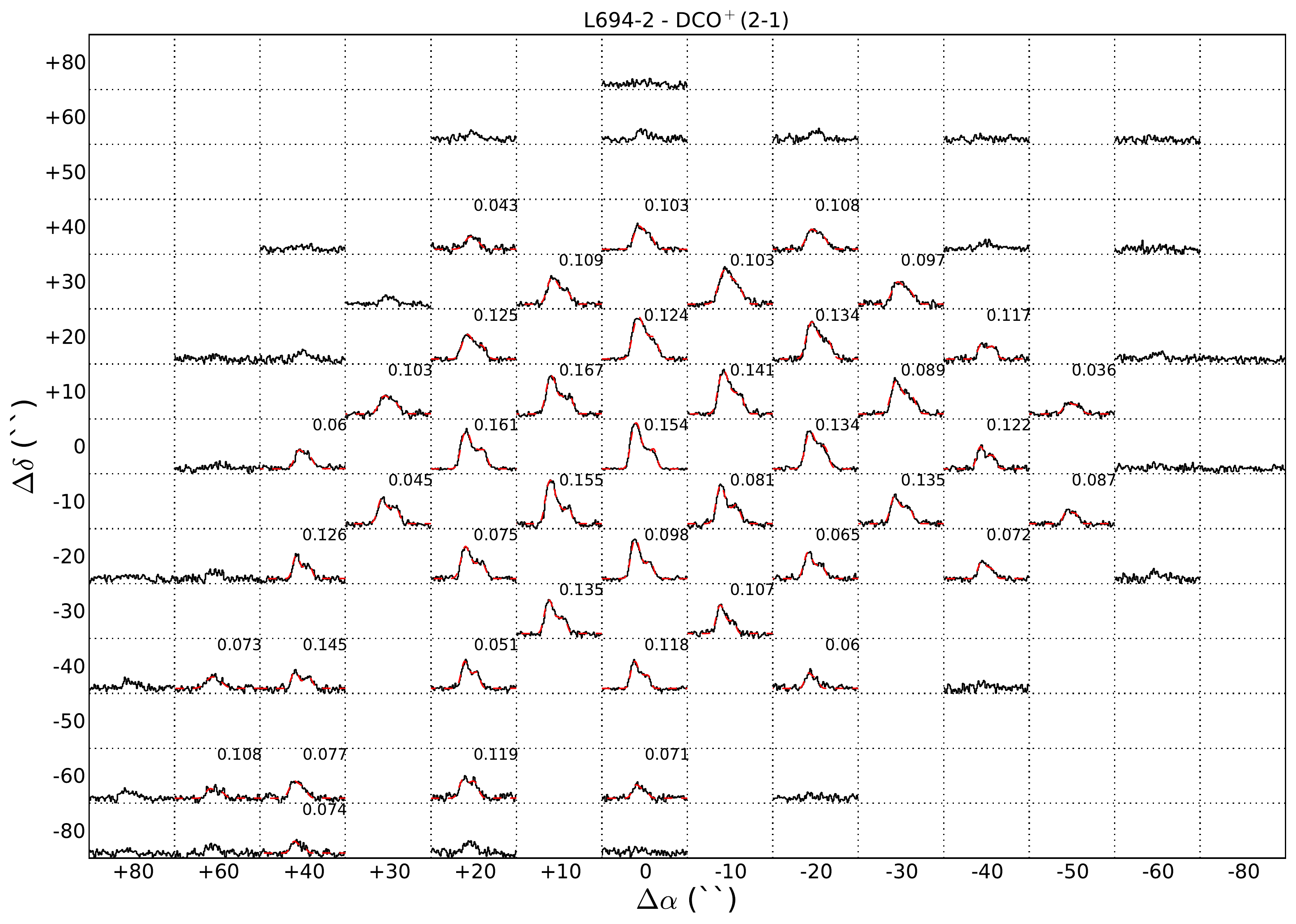}
\caption{Same as Figure 2 for the core L694-2 using the tracer DCO$^+$(2-1).  The map's central pointing is 19:41:04.5, +10:57:02 (J2000).  All spectra are centered on 8.7 $\leq$ V$_{LSR}$ (km s$^{-1}$) $\leq$ 10.7 and -0.5 $\leq$ T$_B$ (K) $\leq$ 5.0.  The two positions with $\Delta\alpha$ = +100 (seen in the south-east of Figure 1) did not have any detections and have been omitted throughout Figure 4.}
\end{sidewaysfigure}

\begin{sidewaysfigure}[ht]
\figurenum{3b}
\includegraphics[scale=0.66]{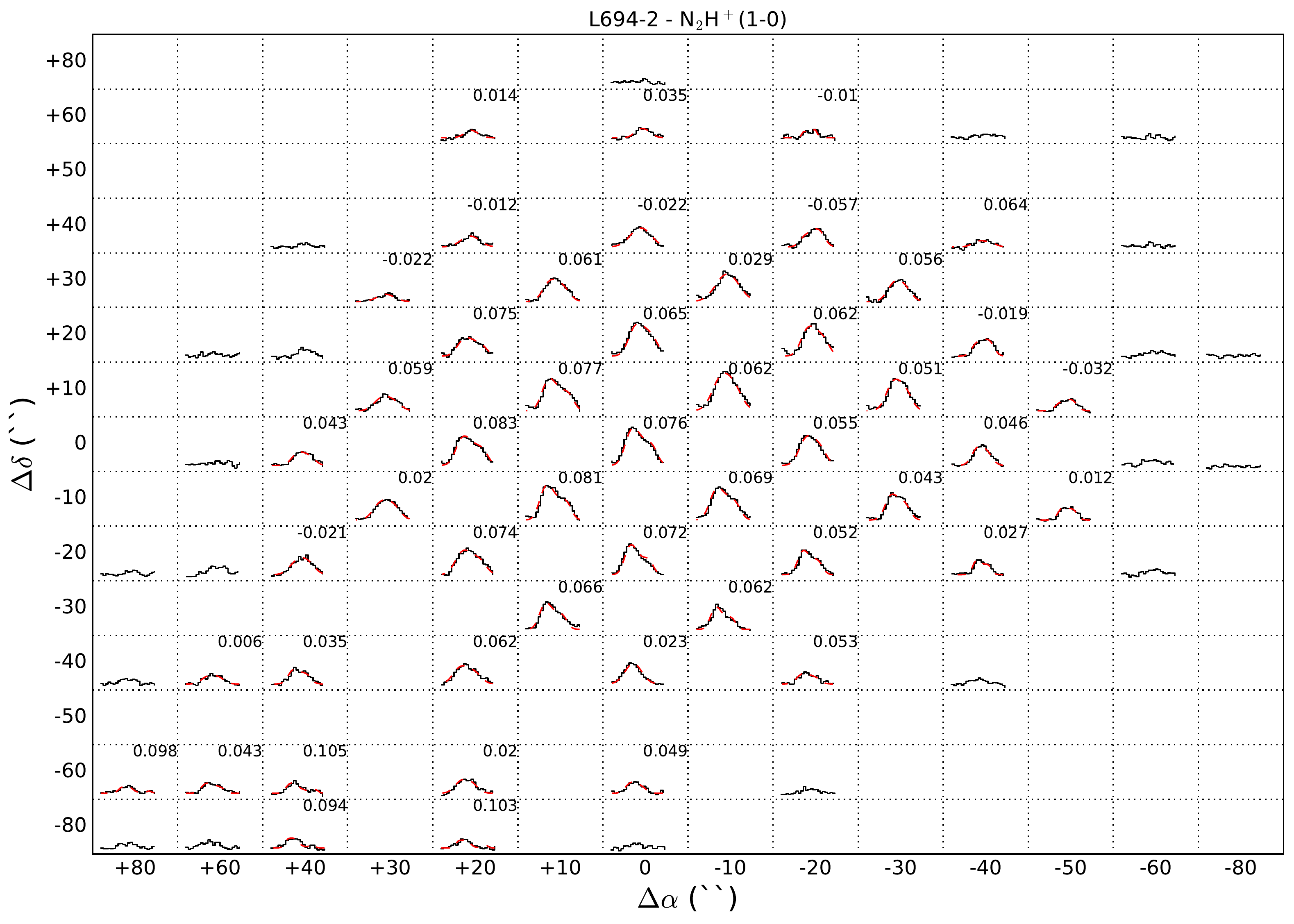}
\caption{Same as Figure 4a using the tracer N$_2$H$^+$(1-0). All spectra are centered on 9.1 $\leq$ V$_{LSR}$ (km s$^{-1}$) $\leq$ 10.2 and -0.5 $\leq$ T$_B$ (K) $\leq$ 4.0.  Only the central hyperfine structure is displayed in each plot.}
\end{sidewaysfigure}

\begin{sidewaysfigure}[ht]
\figurenum{3c}
\includegraphics[scale=0.66]{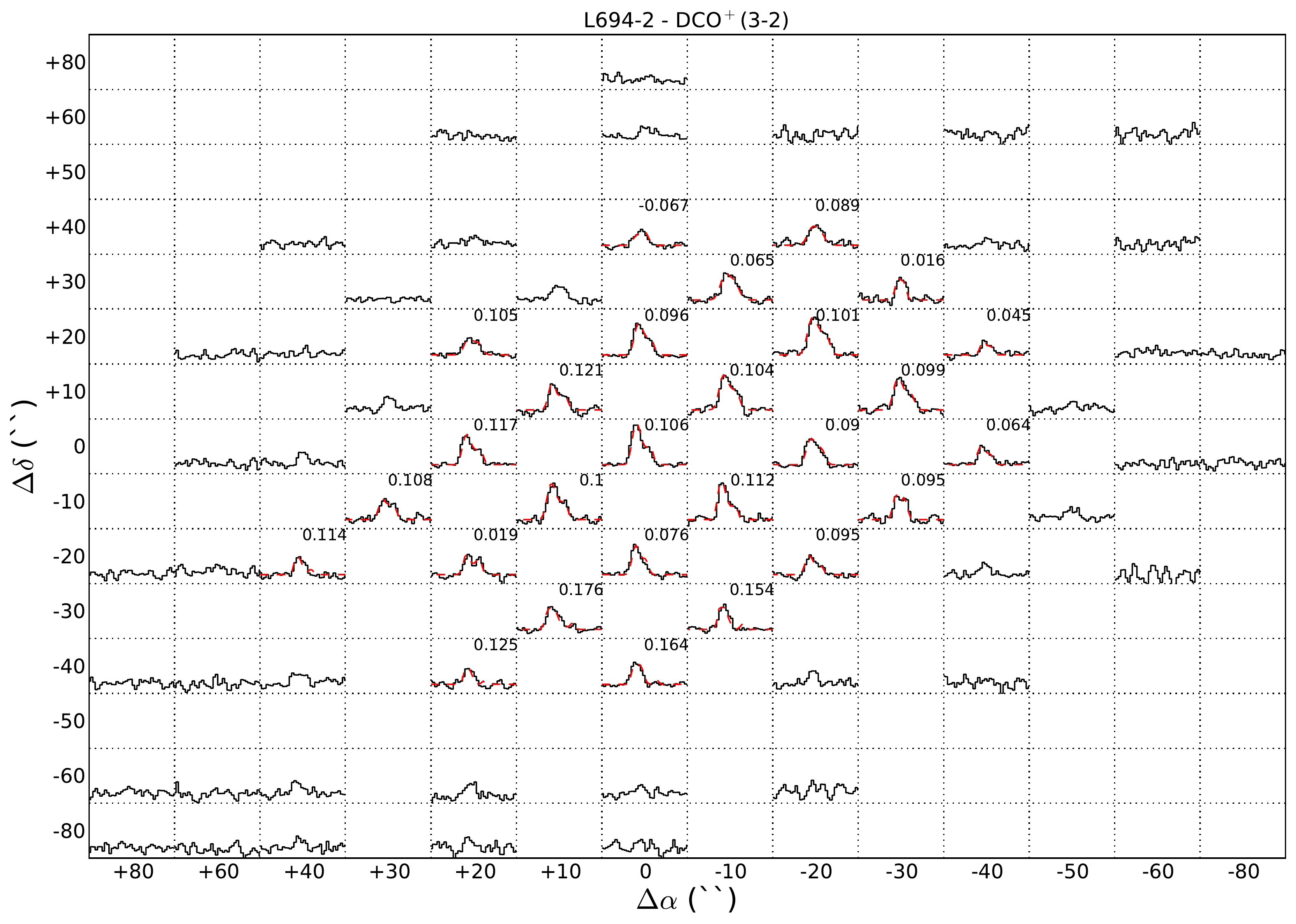}
\caption{Same as Figure 3a for the tracer DCO$^+$(3-2).  All spectra are centered on 8.7 $\leq$ V$_{LSR}$ (km s$^{-1}$) $\leq$ 10.7. and -0.5 $\leq$ T$_B$ (K) $\leq$ 2.5.}
\end{sidewaysfigure}

\begin{sidewaysfigure}[ht]
\figurenum{3d}
\includegraphics[scale=0.66]{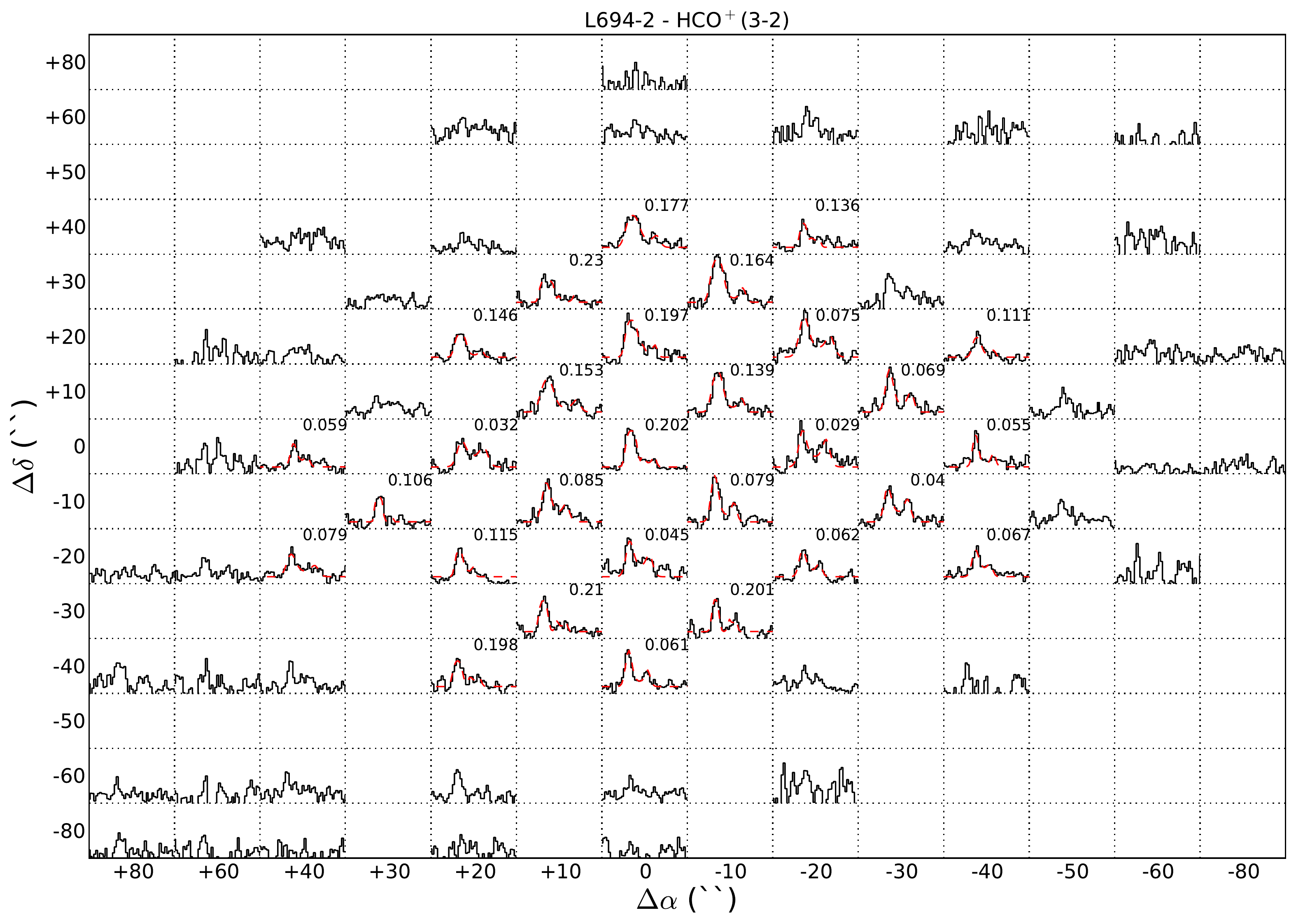}
\caption{Same as Figure 3a for the tracer HCO$^+$(3-2).  All spectra are centered on 8.7 $\leq$ V$_{LSR}$ (km s$^{-1}$) $\leq$ 10.7 and -0.5 $\leq$ T$_B$ (K) $\leq$ 3.5.}
\end{sidewaysfigure}

\begin{sidewaysfigure}[ht]
\figurenum{4a}
\includegraphics[scale=0.66]{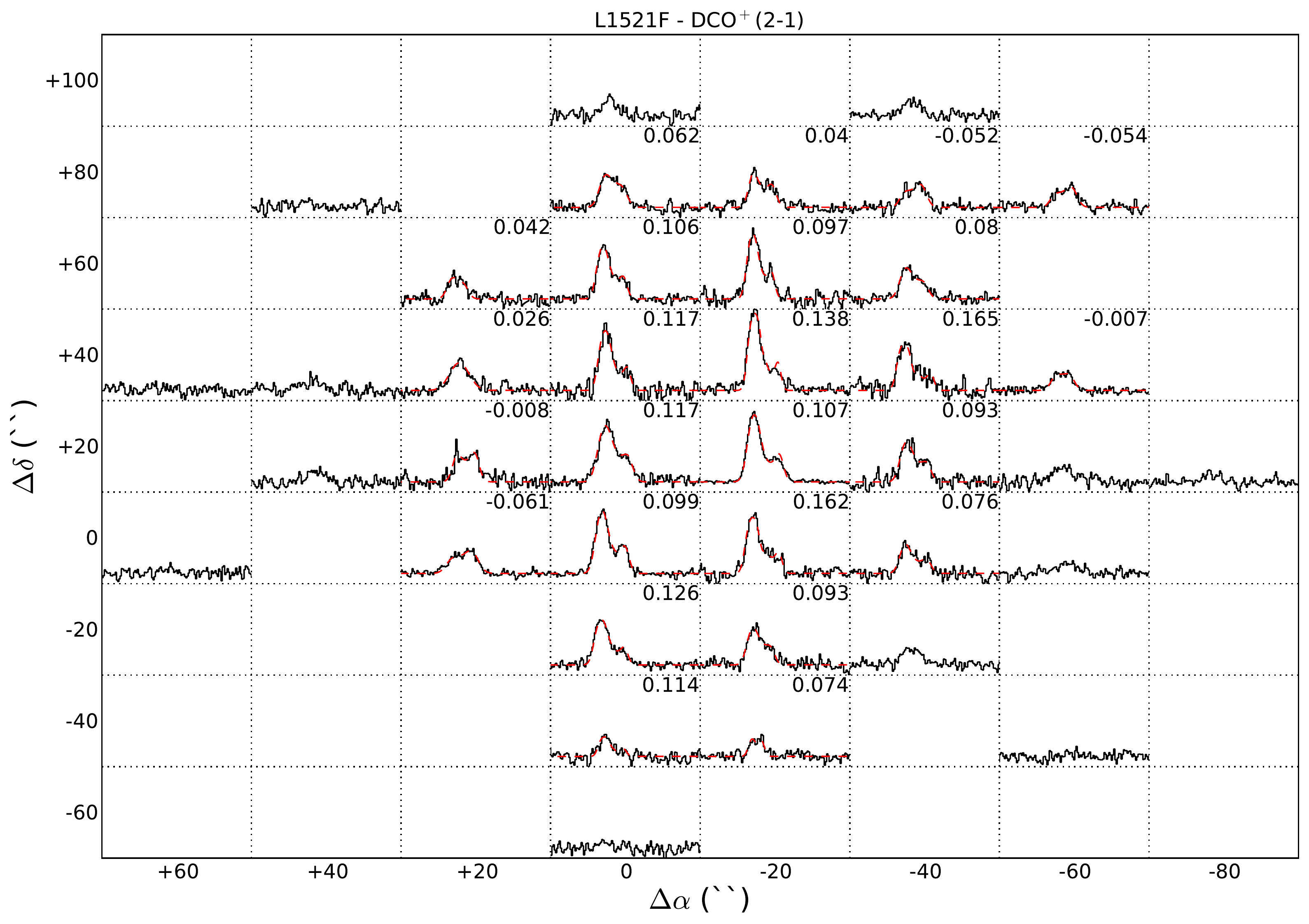}
\caption{Same as Figure 2 for the core L1521F using the tracer DCO$^+$(2-1).  The map's central pointing is 04:28:39.8, +26:51:15 (J2000).  All spectra are centered on 5.5 $\leq$ V$_{LSR}$ (km s$^{-1}$) $\leq$ 8.0 and -0.5 $\leq$ T$_B$ (K) $\leq$ 4.0}
\end{sidewaysfigure}

\begin{sidewaysfigure}[ht]
\figurenum{4b}
\includegraphics[scale=0.66]{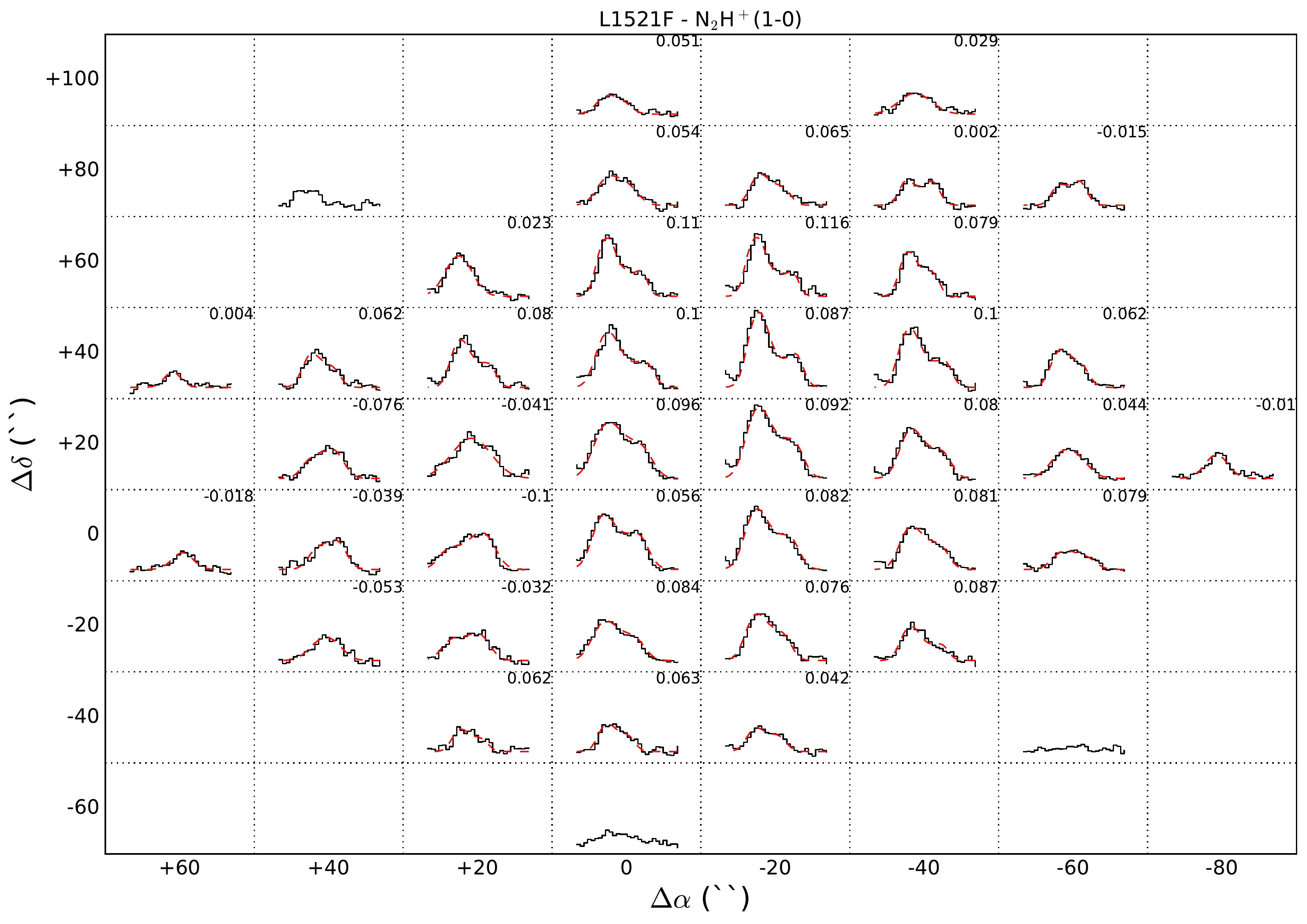}
\caption{Same as Figure 4a using the tracer N$_2$H$^+$(1-0).  All spectra are centered on 5.9 $\leq$ V$_{LSR}$ (km s$^{-1}$) $\leq$ 7.2 and -0.5 $\leq$ T$_B$ (K) $\leq$ 3.4.  Only the central hyperfine structure is displayed in each plot.}
\end{sidewaysfigure}

\begin{sidewaysfigure}[ht]
\figurenum{4c}
\includegraphics[scale=0.66]{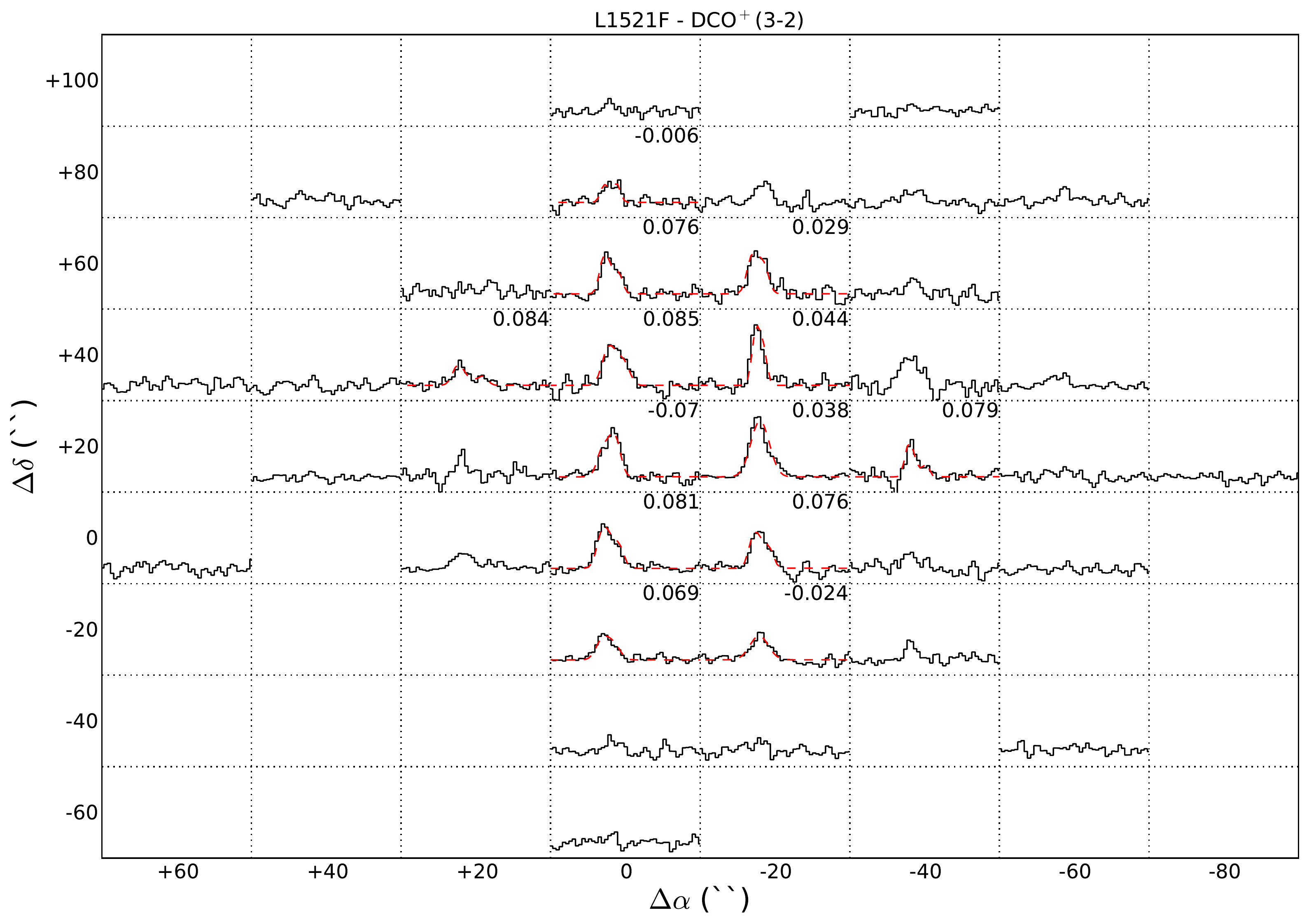}
\caption{Same as Figure 4a for the tracer DCO$^+$(3-2).  All spectra are centered on 5.5 $\leq$ V$_{LSR}$ (km s$^{-1}$) $\leq$ 8.0 and -0.5 $\leq$ T$_B$ (K) $\leq$ 2.5.}
\end{sidewaysfigure}

\begin{sidewaysfigure}[ht]
\figurenum{4d}
\includegraphics[scale=0.66]{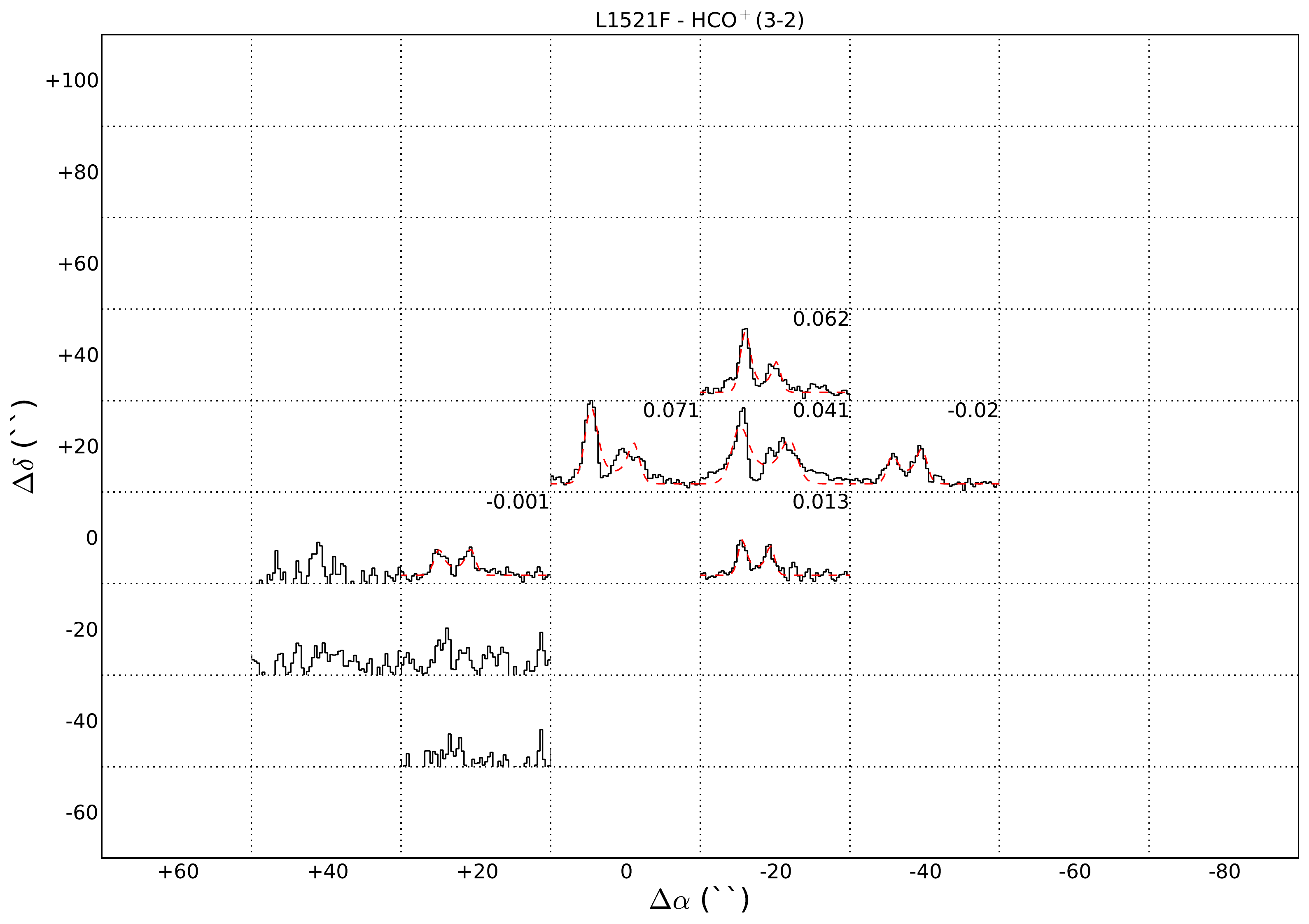}
\caption{Same as Figure 4a for the tracer HCO$^+$(3-2).  All spectra are centered on 5.5 $\leq$ V$_{LSR}$ (km s$^{-1}$) $\leq$ 8.0 and -0.5 $\leq$ T$_B$ (K) $\leq$ 5.0.}
\end{sidewaysfigure}

\begin{figure}[ht]
\figurenum{5}

\epsscale{0.73}
\centering
\plotone{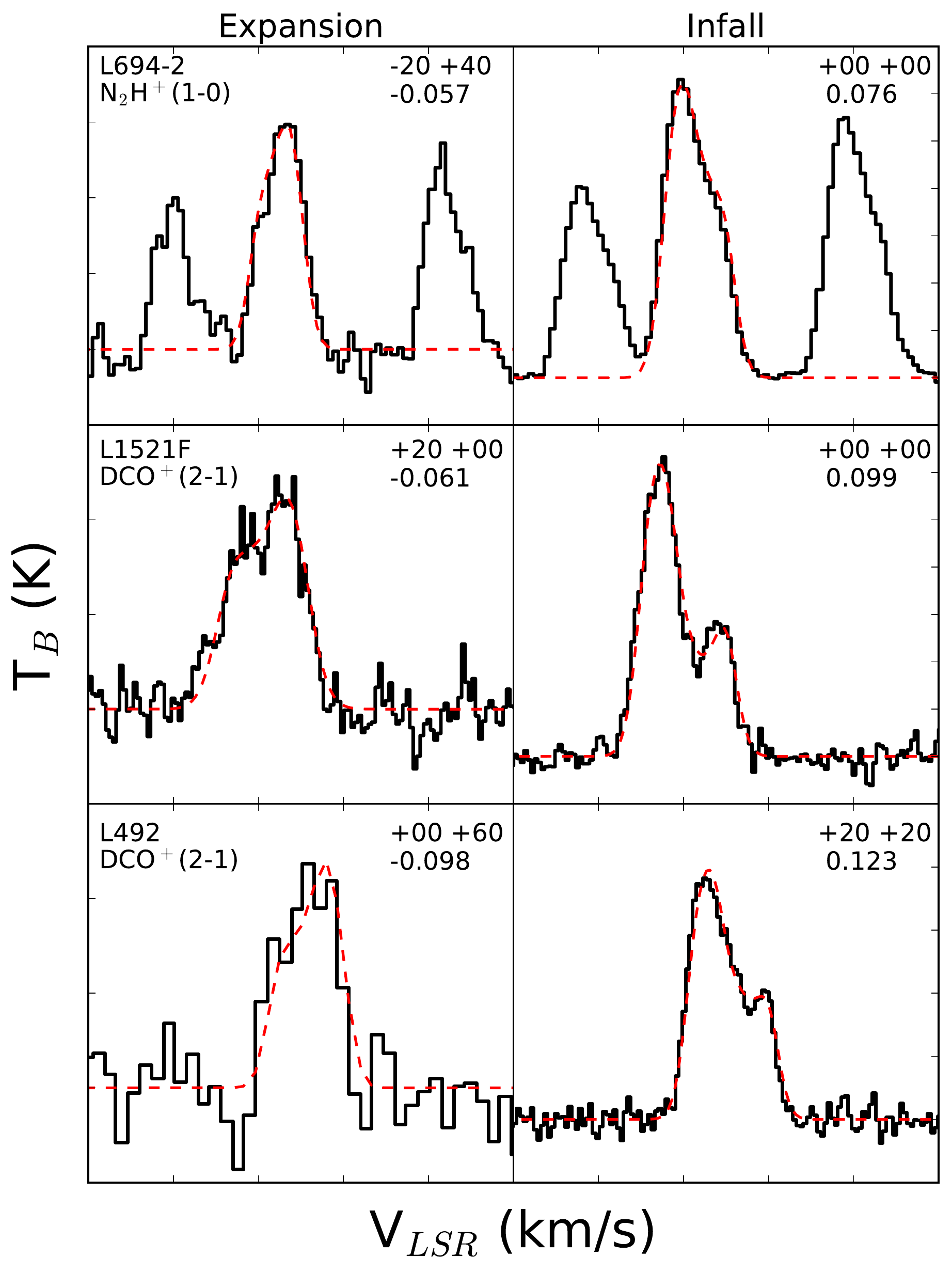}
\caption{Comparison between asymmetrically red spectra indicating expansion (left column) and asymmetrically blue spectra indicating infall (right column) observed in L694-2 (top row), L1521F (middle row), and L492 (bottom row).  Each row shows a spectra observed in a single molecular tracer at a particular pointing within the core indicated by the radial offset in the upper right corner of each subplot.  The observed spectrum is shown in black while the best fit HILL5 model is plotted in red.  The v$_{in}$ estimate in km s$^{-1}$ corresponding to the best fit model is also provided in the upper right corner of each subplot.  Each y-axis tick marker represents an increase of 0.5 K (with -0.5 K marking the bottom of each subplot's y-axis), while each x-axis tick marker represents an increase of 0.5 km s$^{-1}$.  Only the central hyperfine component of the N$_2$H$^+$(1-0) spectra were used for our fitting procedure (see \S 3.2).} 
\end{figure}

\begin{figure}[ht]
\figurenum{6}

\epsscale{0.47}
\plotone{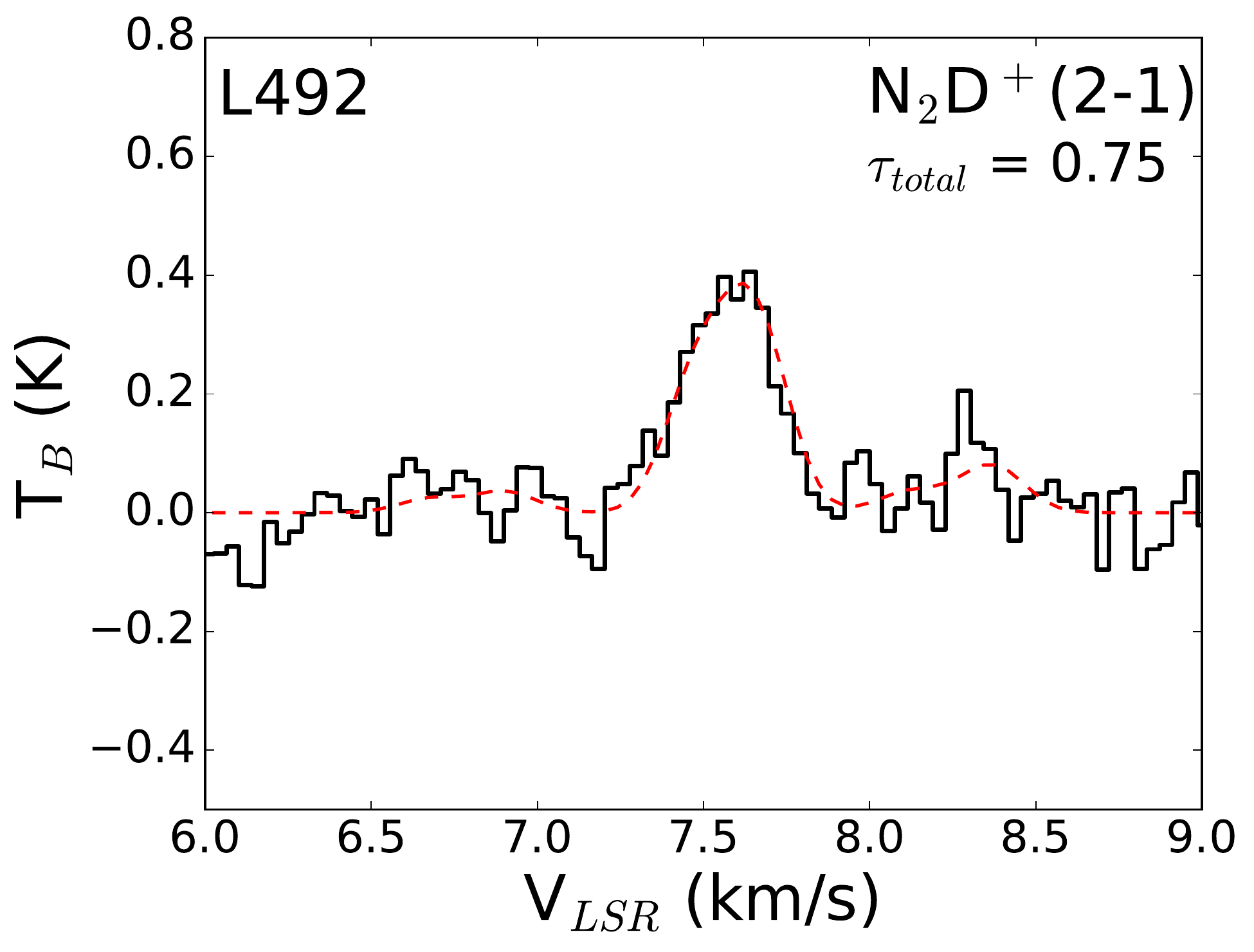} \\
\plotone{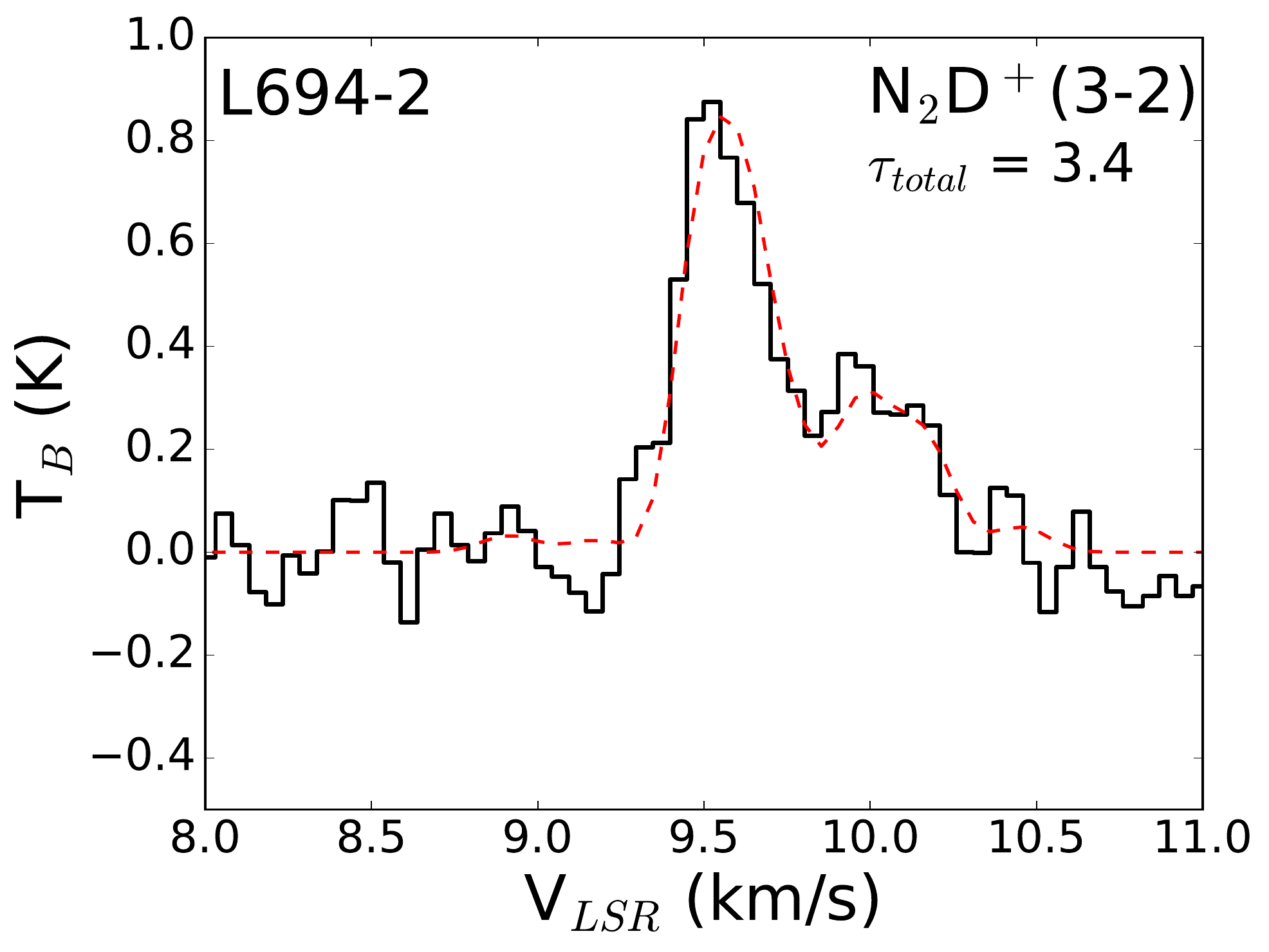} \\
\plotone{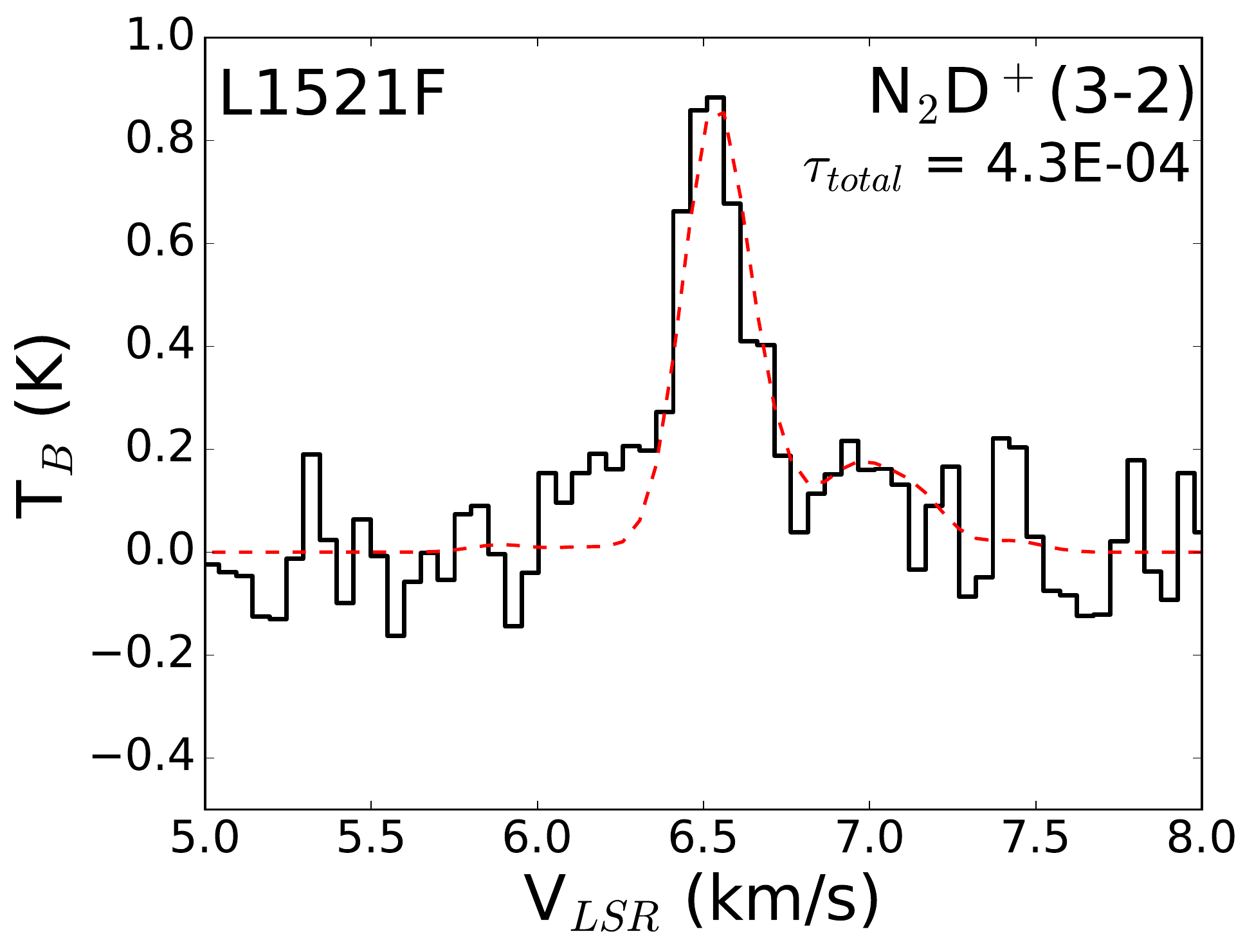}
\caption{Spectra (black) of optically thin N$_2$D$^+$(2-1) and N$_2$D$^+$(3-2) emission toward the dust continuum peak of L492 (top), L694-2 (middle), and L1521F (bottom). A Gaussian fit to the hyperfine structures, based on rest frequencies calculated by \cite{Gerin_2001}, is overplotted in red.  The total optical depth (i.e., the sum of the optical depths of each individual hyperfine component)  found from the Gaussian fitting is displayed in the upper right of each panel.  Although $\tau_{total}$ $>$ 1 for L694-2, all individual hyperfine components are optically thin (see \S 3.1).  F-tests suggest these spectra are better fit by a Gaussian model than the HILL5 model, which may indicate single velocity components along the line of sight (see \S 3.1). Only the central hyperfine components are shown.}
\end{figure}

\begin{figure}[ht]
\figurenum{7}

\epsscale{0.85}
\centering
\plotone{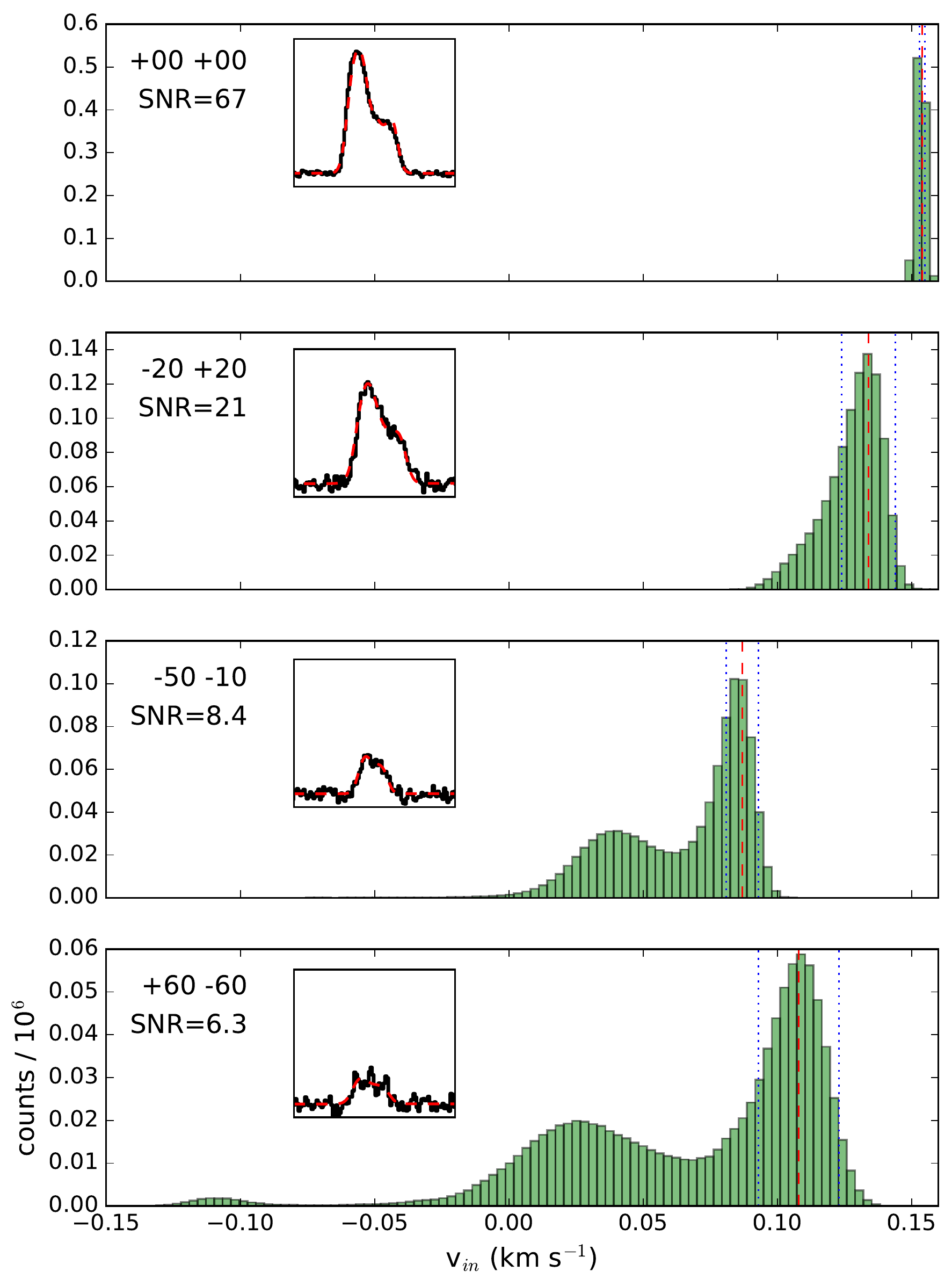}
\caption{Probability mass functions of v$_{in}$ found by MCMC fitting of the HILL5 model to four separate DCO$^+$(2-1) spectra from L694-2 with varying levels of SNR.  The fitted spectrum's radial offset ($\Delta\alpha$, $\Delta\delta$) from the map central pointing and SNR are shown in the upper left corner of each panel.  The inset plot in each panel displays the observed spectrum (black) along with its best fit HILL5 model (red) found independently by the MPFIT non-linear least squares curve fitting routine.  The red dashed line and blue dotted lines in the main plots show the best fit v$_{in}$ estimate and corresponding 1-$\sigma$ uncertainties, respectively, found by MPFIT.}
\end{figure}

\begin{figure}[ht]
\figurenum{8}

\epsscale{1.0}
\centering
\plotone{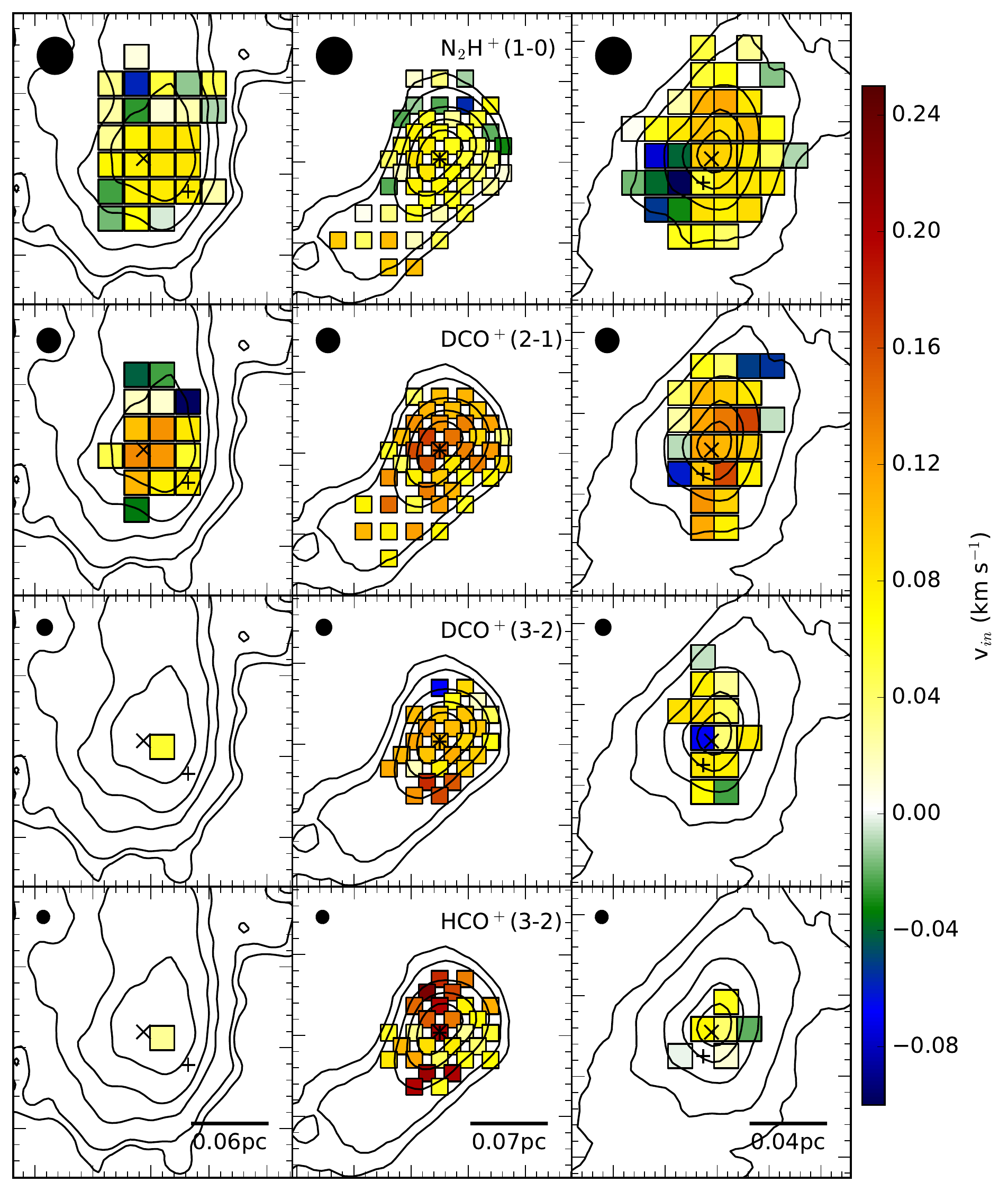}
\caption{Infall/expansion velocity maps for L492 (left column), L694-2 (middle column), and L1521F (right column) using only spectra with SNR $>$ 6. Each row represents a single molecular tracer; from top to bottom: N$_2$H$^+$(1-0), DCO$^+$(2-1), DCO$^+$(3-2), and HCO$^+$(3-2).  Velocities are in km~s$^{-1}$ with positive velocities representing infall and negative velocities representing expansion.  FWHM beam sizes are shown in the upper left corners.  Background contours are the same as in Figure 1.  The black ``x" in each map denotes the position of the core's dust continuum peak, while the ``+" marks the $\Delta\alpha$ = 0, $\Delta\delta$ = 0 position.}
\end{figure}

\begin{figure}[ht]
\figurenum{9}

\epsscale{0.9}
\centering
\plotone{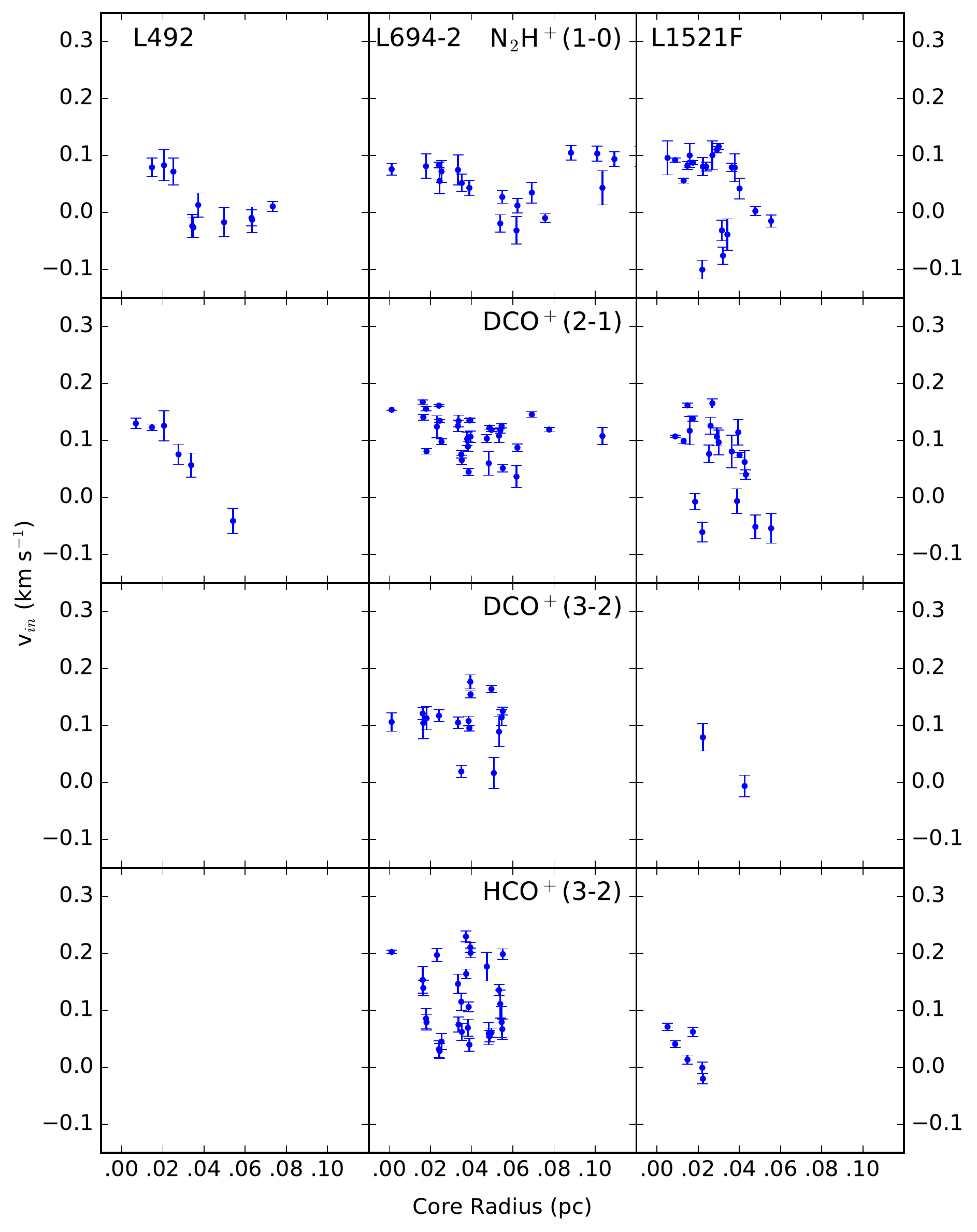}
\caption{Infall velocity (v$_{in}$) versus core radius from dust continuum peak for L492 (left column), L694-2 (middle column), and L1521F (right column) using pointings with SNR $>$ 6 and v$_{in}$ uncertainties less than 0.03 km s$^{-1}$.  All x- and y-axes are uniform.  Each row represents a single molecular tracer; from top to bottom: N$_2$H$^+$(1-0), DCO$^+$(2-1), DCO$^+$(3-2), and HCO$^+$(3-2).  The distances to each core as listed in Table 1 were used to convert the radial offsets from the dust continuum peak for each pointing into a physical distance in parsecs.}
\end{figure}

\clearpage
\begin{deluxetable}{ccccccc}
\tabletypesize{\tiny}
\tablewidth{0pt}
\tablecolumns{7}
\tablecaption{Summary of Properties from Literature\label{Properties}}
\tablehead{\colhead{Core} & \colhead{Distance\tablenotemark{a}} & \colhead{Radius} & \colhead{Radius} & \colhead{Mass} & \colhead{N(N$_2$D$^+$)/N(N$_2$H$^+$)\tablenotemark{a}} & \colhead{HCO$^+$(3-2) V$_{in}$\tablenotemark{b}} \\
 & pc & \arcsec & pc & M$_\odot$ &  & km~s$^{-1}$}
\startdata
L492 & 200 & 161.1\tablenotemark{c} & 0.156 & 5.2\tablenotemark{c} & 0.05 $\pm$ 0.01 & static\tablenotemark{e}\\
L694-2 & 250 & 66.4\tablenotemark{d} & 0.080 & 6.8\tablenotemark{d} & 0.26 $\pm$ 0.05 & 0.16 $\pm$ 0.01 \\ 
L1521F & 140 & 83.0\tablenotemark{d} & 0.056 & 4.4\tablenotemark{d} & 0.10 $\pm$ 0.02 & 0.42 $\pm$ 0.01 \\ 
\enddata
\tablenotetext{a}{\citep{2005ApJ...619..379C}}
\tablenotetext{b}{\citep{schnee_2013}}
\tablenotetext{c}{\citep{kauffmann_2008}}
\tablenotetext{d}{\citep{0067-0049-175-1-277}}
\tablenotetext{e}{Although \cite{schnee_2013} labeled L492 as a ``static core" due to the subtlety of its infall asymmetry in HCO$^+$(3-2), this core has been classified as a ``contracting core" by \cite{2007ApJ...664..928S} due to infall asymmetries in all three hyperfine components of HCN(1-0).  \cite{2011ApJ...734...60L} also classify L492 as a ``contracting core" based on infall asymmetries determined from CS, HCN, and N$_2$H$^+$ spectra.}

\end{deluxetable}

\begin{deluxetable}{ccccccc}
\tabletypesize{\tiny}
\tablewidth{0pt}
\tablecolumns{7}
\tablecaption{Summary of IRAM Observations\label{IRAM}}
\tablehead{\colhead{Molecule} & \colhead{Frequency} & \colhead{A$_{ul}$\tablenotemark{a}} & \colhead{$\gamma_{ul}$\tablenotemark{b}} & \colhead{n$_{crit}$\tablenotemark{c}} & \colhead{Angular Resolution} & \colhead{Spectral Resolution}\\
 & GHz & s$^{-1}$ & cm$^{-3}$ s$^{-1}$ & cm$^{-3}$ & \arcsec & km~s$^{-1}$}
\startdata
N$_2$H$^+$(1-0) & 93.1740 & 3.6280E-5 & 2.6E-10 & 1.4E+5 & 27.0 & 0.031
\\ 
DCO$^+$(2-1) & 144.077 & 2.1358E-4 & 3.8E-10 & 5.6E+5 & 17.5 & 0.020
\\ 
DCO$^+$(3-2) & 216.113 & 7.7217E-4 & 4.3E-10 & 1.8E+6 & 11.6 & 0.054
\\ 
HCO$^+$(3-2) & 267.558 & 1.4757E-3 & 4.3E-10 & 3.4E+6 & 9.41 & 0.044
\\ 
\enddata
\tablenotetext{a}{Einstein A-coefficient for given transition, as listed in the Leiden Atomic and Molecular Database (LAMDA) \citep{Schoier_2005}}
\tablenotetext{b}{Collisional rate coefficient for given transition at 10$~$K for collisions with H$_2$, as listed in LAMDA \citep{Schoier_2005, Flower_1999}}
\tablenotetext{c}{n$_{crit}$(u-l) = A$_{ul}$ / $\gamma_{ul}$}
\end{deluxetable}

\begin{deluxetable}{ccccccccccc}
\tabletypesize{\tiny}
\tablewidth{0pt}
\tablecolumns{11}
\tablecaption{L492 Infall/Expansion Measurements\label{L492}}
\tablehead{\colhead{No.} & \colhead{$\Delta\alpha$\tablenotemark{a}} & \colhead{$\Delta\delta$\tablenotemark{a}} & \colhead{N$_2$H$^+$(1-0)\tablenotemark{b}} & \colhead{N$_2$H$^+$(1-0)\tablenotemark{c}} & \colhead{DCO$^+$(2-1)\tablenotemark{b}} & \colhead{DCO$^+$(2-1)\tablenotemark{c}} & \colhead{DCO$^+$(3-2)\tablenotemark{b}}  & \colhead{DCO$^+$(3-2)\tablenotemark{c}}  & \colhead{HCO$^+$(3-2)\tablenotemark{b}} & \colhead{HCO$^+$(3-2)\tablenotemark{c}}\\
 & \arcsec & \arcsec & v$_{in}$ (km~s$^{-1}$) & SNR & v$_{in}$ (km~s$^{-1}$) & SNR & v$_{in}$ (km~s$^{-1}$) & SNR & v$_{in}$ (km~s$^{-1}$) & SNR}
\startdata
1 & -20 & -20 & 0.006 $\pm$ 0.008 & 3.2 & -0.030 $\pm$ 0.051 & 3.0 & -0.037 $\pm$ 0.078 & 3.0 & 0.175 $\pm$ 0.015 & 3.6 \\2 & -20 & 0 & 0.024 $\pm$ 2.362 & 8.4 & -0.133 $\pm$ 0.022 & 3.6 & -0.002 $\pm$ 0.023 & 3.6 & -0.184 $\pm$ 0.020 & 2.6 \\3 & -20 & 20 & 0.004 $\pm$ 0.021 & 5.0 & -0.082 $\pm$ 1.853 & 3.8 & \nodata & \nodata & \nodata & \nodata \\4 & -20 & 40 & 0.046 $\pm$ 0.017 & 5.5 & -0.211 $\pm$ 0.053 & 3.4 & \nodata & \nodata & \nodata & \nodata \\5 & -20 & 60 & -0.010 $\pm$ 0.014 & 6.5 & -0.250 $\pm$ 0.000 & 2.5 & -0.024 $\pm$ 0.028 & 2.7 & -0.156 $\pm$ 0.110 & 3.3 \\6 & -20 & 80 & 0.049 $\pm$ 0.197 & 6.3 & -0.088 $\pm$ 0.015 & 3.1 & 0.156 $\pm$ 0.000 & 2.2 & -0.038 $\pm$ 0.175 & 3.5 \\7 & 0 & -40 & -0.113 $\pm$ 0.022 & 4.1 & 0.073 $\pm$ 0.000 & 2.2 & 0.247 $\pm$ 0.000 & 3.0 & 0.047 $\pm$ 0.018 & 2.5 \\8 & 0 & -20 & 0.018 $\pm$ 0.024 & 5.8 & 0.083 $\pm$ 0.021 & 3.5 & -0.117 $\pm$ 0.092 & 3.4 & \nodata & \nodata \\9 & 0 & 0 & 0.078 $\pm$ 0.052 & 25.6 & 0.078 $\pm$ 0.264 & 9.6 & 0.082 $\pm$ 0.139 & 3.2 & 0.047 $\pm$ 0.112 & 2.9 \\10 & 0 & 20 & 0.081 $\pm$ 0.037 & 28.0 & 0.057 $\pm$ 0.021 & 8.4 & -0.182 $\pm$ 0.026 & 2.5 & \nodata & \nodata \\11 & 0 & 40 & 0.081 $\pm$ 0.076 & 19.6 & 0.080 $\pm$ 0.080 & 6.5 & 0.062 $\pm$ 0.044 & 3.5 & \nodata & \nodata \\12 & 0 & 60 & 0.023 $\pm$ 0.064 & 12.8 & -0.098 $\pm$ 0.032 & 6.6 & -0.018 $\pm$ 0.020 & 3.3 & \nodata & \nodata \\13 & 0 & 80 & -0.013 $\pm$ 0.023 & 6.6 & -0.170 $\pm$ 0.055 & 5.1 & \nodata & \nodata & \nodata & \nodata \\14 & 0 & 100 & -0.107 $\pm$ 0.033 & 3.3 & 0.000 $\pm$ 0.053 & 2.7 & 0.536 $\pm$ 0.000 & 3.3 & \nodata & \nodata \\15 & 20 & -40 & -0.077 $\pm$ 0.061 & 5.7 & 0.161 $\pm$ 0.033 & 3.3 & \nodata & \nodata & \nodata & \nodata \\16 & 20 & -20 & -0.004 $\pm$ 1.423 & 8.0 & 0.130 $\pm$ 0.038 & 3.1 & 0.213 $\pm$ 0.154 & 3.7 & \nodata & \nodata \\17 & 20 & 0 & 0.079 $\pm$ 0.042 & 31.6 & 0.075 $\pm$ 0.018 & 10.8 & 0.058 $\pm$ 0.061 & 2.8 & \nodata & \nodata \\18 & 20 & 20 & 0.079 $\pm$ 0.016 & 63.0 & 0.123 $\pm$ 0.006 & 20.6 & 0.055 $\pm$ 0.137 & 9.1 & 0.029 $\pm$ 0.500 & 6.6 \\19 & 20 & 40 & 0.083 $\pm$ 0.027 & 42.4 & 0.126 $\pm$ 0.026 & 17.3 & 0.009 $\pm$ 0.013 & 4.9 & \nodata & \nodata \\20 & 20 & 60 & 0.013 $\pm$ 0.021 & 14.5 & 0.014 $\pm$ 0.039 & 7.8 & 0.114 $\pm$ 0.012 & 3.5 & \nodata & \nodata \\21 & 20 & 80 & 0.052 $\pm$ 0.111 & 9.5 & -0.024 $\pm$ 0.121 & 7.2 & \nodata & \nodata & \nodata & \nodata \\22 & 20 & 100 & -0.083 $\pm$ 0.243 & 5.5 & 0.016 $\pm$ 0.023 & 3.6 & \nodata & \nodata & \nodata & \nodata \\23 & 40 & -40 & 0.095 $\pm$ 0.148 & 4.2 & -0.070 $\pm$ 0.050 & 2.6 & -0.300 $\pm$ 0.000 & 3.5 & 0.146 $\pm$ 0.324 & 3.3 \\24 & 40 & -20 & 0.067 $\pm$ 0.190 & 10.3 & -0.034 $\pm$ 0.565 & 7.2 & \nodata & \nodata & \nodata & \nodata \\25 & 40 & 0 & 0.076 $\pm$ 0.160 & 12.3 & 0.107 $\pm$ 0.118 & 6.1 & 0.002 $\pm$ 0.024 & 3.5 & \nodata & \nodata \\26 & 40 & 20 & 0.076 $\pm$ 0.030 & 36.0 & 0.130 $\pm$ 0.009 & 10.4 & 0.075 $\pm$ 0.202 & 4.5 & \nodata & \nodata \\27 & 40 & 40 & 0.074 $\pm$ 0.059 & 12.4 & 0.102 $\pm$ 0.275 & 6.6 & -0.298 $\pm$ 0.038 & 2.3 & \nodata & \nodata \\28 & 40 & 60 & -0.026 $\pm$ 0.017 & 12.7 & 0.018 $\pm$ 0.034 & 6.1 & -0.160 $\pm$ 0.018 & 4.6 & \nodata & \nodata \\29 & 40 & 80 & -0.059 $\pm$ 0.053 & 9.5 & -0.041 $\pm$ 0.022 & 7.0 & \nodata & \nodata & \nodata & \nodata \\30 & 40 & 100 & 0.011 $\pm$ 0.009 & 6.0 & 0.114 $\pm$ 0.018 & 2.7 & 0.423 $\pm$ 0.059 & 2.8 & 0.104 $\pm$ 0.132 & 2.4 \\31 & 60 & -40 & -0.008 $\pm$ 0.020 & 3.7 & -0.171 $\pm$ 0.067 & 2.7 & 0.049 $\pm$ 0.062 & 3.1 & -0.086 $\pm$ 0.027 & 3.9 \\32 & 60 & -20 & -0.017 $\pm$ 0.026 & 6.0 & 0.175 $\pm$ 0.024 & 3.5 & -0.300 $\pm$ 0.000 & 3.1 & -0.104 $\pm$ 0.027 & 4.0 \\33 & 60 & 0 & -0.024 $\pm$ 0.020 & 9.4 & 0.151 $\pm$ 0.011 & 5.9 & \nodata & \nodata & \nodata & \nodata \\34 & 60 & 20 & 0.072 $\pm$ 0.024 & 15.7 & 0.049 $\pm$ 0.048 & 6.5 & -0.024 $\pm$ 0.037 & 2.6 & \nodata & \nodata \\35 & 60 & 40 & 0.030 $\pm$ 1.787 & 11.3 & 0.084 $\pm$ 0.456 & 5.3 & -0.113 $\pm$ 0.032 & 2.7 & \nodata & \nodata \\36 & 60 & 60 & 0.025 $\pm$ 1.013 & 9.2 & -0.172 $\pm$ 0.044 & 3.5 & -0.300 $\pm$ 0.180 & 2.6 & \nodata & \nodata \\37 & 60 & 80 & 0.032 $\pm$ 0.044 & 9.0 & -0.054 $\pm$ 0.069 & 3.5 & 0.009 $\pm$ 0.023 & 3.1 & -0.135 $\pm$ 0.014 & 2.9 \\38 & 80 & -20 & -0.062 $\pm$ 0.465 & 4.0 & -0.218 $\pm$ 0.228 & 2.8 & 0.062 $\pm$ 0.063 & 3.2 & 0.016 $\pm$ 0.000 & 2.8 \\39 & 80 & 0 & -0.055 $\pm$ 0.023 & 4.9 & 0.402 $\pm$ 0.058 & 2.6 & -0.135 $\pm$ 0.089 & 3.6 & 0.162 $\pm$ 0.406 & 3.6 \\40 & 80 & 20 & 0.060 $\pm$ 0.024 & 5.9 & -0.257 $\pm$ 0.068 & 3.2 & \nodata & \nodata & \nodata & \nodata \\41 & 80 & 40 & -0.143 $\pm$ 0.044 & 5.4 & -0.242 $\pm$ 0.062 & 2.6 & \nodata & \nodata & \nodata & \nodata \\42 & 80 & 60 & -0.063 $\pm$ 0.022 & 2.9 & 0.108 $\pm$ 0.036 & 3.3 & 0.011 $\pm$ 0.030 & 3.3 & 0.277 $\pm$ 0.525 & 2.8 \\

\enddata

\tablenotetext{a}{Offset from map's central pointing: 18:15:46.08, -03:46:12.8 (J2000)}
\tablenotetext{b}{Measured infall (positive) or expansion (negative) velocity}
\tablenotetext{c}{SNR of the spectrum}
\tablecomments{A zero $\pm$ value indicates a low SNR spectrum for which MPFIT was unable to calculate an uncertainty (these sources were omitted from all analyses presented in this paper).}
\end{deluxetable}

\begin{deluxetable}{ccccccccccc}
\tabletypesize{\tiny}
\tablewidth{0pt}
\tablecolumns{11}
\tablecaption{L694-2 Infall/Expansion Measurements\label{L694-2}}
\tablehead{\colhead{No.} & \colhead{$\Delta\alpha$} & \colhead{$\Delta\delta$} & \colhead{N$_2$H$^+$(1-0)} & \colhead{N$_2$H$^+$(1-0)} & \colhead{DCO$^+$(2-1)} & \colhead{DCO$^+$(2-1)} & \colhead{DCO$^+$(3-2)}  & \colhead{DCO$^+$(3-2)}  & \colhead{HCO$^+$(3-2)} & \colhead{HCO$^+$(3-2)}\\
 & \arcsec & \arcsec & v$_{in}$ (km~s$^{-1}$) & SNR & v$_{in}$ (km~s$^{-1}$) & SNR & v$_{in}$ (km~s$^{-1}$) & SNR & v$_{in}$ (km~s$^{-1}$) & SNR}
\startdata
1 & 0 & 80 & 0.079 $\pm$ 0.011 & 3.4 & -0.128 $\pm$ 0.330 & 2.9 & 0.054 $\pm$ 0.001 & 2.7 & 0.042 $\pm$ 0.025 & 3.1 \\2 & 20 & 60 & 0.014 $\pm$ 1.122 & 6.0 & -0.183 $\pm$ 0.013 & 4.2 & 0.042 $\pm$ 0.117 & 2.6 & 0.139 $\pm$ 0.170 & 2.5 \\3 & 0 & 60 & 0.035 $\pm$ 0.133 & 6.3 & 0.116 $\pm$ 0.042 & 5.4 & 0.094 $\pm$ 0.098 & 3.9 & 0.046 $\pm$ 0.021 & 4.8 \\4 & -20 & 60 & -0.010 $\pm$ 0.008 & 6.4 & -0.016 $\pm$ 0.020 & 5.2 & -0.002 $\pm$ 0.027 & 2.8 & 0.025 $\pm$ 0.020 & 4.1 \\5 & -40 & 60 & 0.016 $\pm$ 0.136 & 4.3 & 0.108 $\pm$ 0.019 & 3.7 & 0.162 $\pm$ 0.000 & 2.8 & -0.005 $\pm$ 0.058 & 3.7 \\6 & -60 & 60 & 0.061 $\pm$ 0.029 & 3.6 & 0.142 $\pm$ 0.000 & 2.9 & 0.071 $\pm$ 0.000 & 3.3 & 0.245 $\pm$ 0.525 & 2.8 \\7 & 40 & 40 & -0.004 $\pm$ 0.013 & 4.3 & -0.041 $\pm$ 0.069 & 2.9 & -0.010 $\pm$ 0.043 & 2.7 & 0.253 $\pm$ 0.026 & 2.9 \\8 & 20 & 40 & -0.012 $\pm$ 1.556 & 9.3 & 0.043 $\pm$ 0.121 & 6.5 & 0.059 $\pm$ 6.422 & 3.3 & 0.135 $\pm$ 0.028 & 4.2 \\9 & 0 & 40 & -0.022 $\pm$ 0.593 & 19.7 & 0.103 $\pm$ 0.007 & 19.1 & -0.067 $\pm$ 0.170 & 8.4 & 0.177 $\pm$ 0.025 & 10.6 \\10 & -20 & 40 & -0.057 $\pm$ 0.098 & 10.9 & 0.108 $\pm$ 0.012 & 11.6 & 0.089 $\pm$ 0.026 & 8.2 & 0.136 $\pm$ 0.010 & 8.2 \\11 & -40 & 40 & 0.064 $\pm$ 0.051 & 6.1 & 0.028 $\pm$ 3.163 & 4.9 & 0.075 $\pm$ 0.085 & 3.3 & 0.184 $\pm$ 0.022 & 4.6 \\12 & -60 & 40 & 0.028 $\pm$ 0.017 & 3.6 & -0.004 $\pm$ 0.024 & 3.7 & 0.112 $\pm$ 0.098 & 2.8 & 0.093 $\pm$ 0.015 & 3.4 \\13 & 30 & 30 & -0.022 $\pm$ 2.484 & 6.7 & 0.029 $\pm$ 0.021 & 5.6 & 0.157 $\pm$ 0.000 & 2.9 & 0.043 $\pm$ 0.081 & 2.2 \\14 & 10 & 30 & 0.061 $\pm$ 0.107 & 17.9 & 0.109 $\pm$ 0.049 & 17.3 & 0.100 $\pm$ 0.014 & 5.7 & 0.230 $\pm$ 0.009 & 8.3 \\15 & -10 & 30 & 0.029 $\pm$ 0.310 & 23.5 & 0.103 $\pm$ 0.051 & 23.5 & 0.065 $\pm$ 0.056 & 10.9 & 0.164 $\pm$ 0.008 & 16.4 \\16 & -30 & 30 & 0.056 $\pm$ 0.070 & 15.0 & 0.097 $\pm$ 0.085 & 10.0 & 0.016 $\pm$ 0.027 & 6.8 & 0.056 $\pm$ 0.023 & 4.6 \\17 & 60 & 20 & 0.009 $\pm$ 0.023 & 3.6 & -0.070 $\pm$ 0.012 & 2.8 & 0.107 $\pm$ 0.055 & 3.0 & 0.007 $\pm$ 0.012 & 3.9 \\18 & 40 & 20 & 0.027 $\pm$ 0.023 & 5.3 & -0.152 $\pm$ 0.016 & 4.8 & 0.167 $\pm$ 0.000 & 3.3 & 0.131 $\pm$ 0.025 & 2.8 \\19 & 20 & 20 & 0.075 $\pm$ 0.026 & 15.7 & 0.125 $\pm$ 0.010 & 18.2 & 0.105 $\pm$ 0.010 & 9.7 & 0.146 $\pm$ 0.017 & 6.8 \\20 & 0 & 20 & 0.065 $\pm$ 0.032 & 50.0 & 0.124 $\pm$ 0.019 & 31.6 & 0.096 $\pm$ 0.044 & 18.4 & 0.197 $\pm$ 0.011 & 9.1 \\21 & -20 & 20 & 0.062 $\pm$ 0.085 & 21.9 & 0.134 $\pm$ 0.010 & 21.6 & 0.101 $\pm$ 0.092 & 17.1 & 0.075 $\pm$ 0.013 & 11.0 \\22 & -40 & 20 & -0.019 $\pm$ 0.015 & 12.4 & 0.117 $\pm$ 0.004 & 8.8 & 0.045 $\pm$ 0.032 & 6.1 & 0.111 $\pm$ 0.025 & 6.9 \\23 & -60 & 20 & -0.005 $\pm$ 0.065 & 4.7 & -0.124 $\pm$ 0.026 & 3.7 & 0.418 $\pm$ 0.071 & 3.0 & 0.105 $\pm$ 0.030 & 3.0 \\24 & -80 & 20 & -0.017 $\pm$ 0.027 & 3.5 & 0.164 $\pm$ 0.032 & 3.2 & 0.204 $\pm$ 0.000 & 2.4 & -0.014 $\pm$ 5.773 & 2.4 \\25 & 30 & 10 & 0.059 $\pm$ 0.187 & 12.8 & 0.103 $\pm$ 0.012 & 11.0 & 0.072 $\pm$ 0.010 & 5.7 & 0.191 $\pm$ 0.023 & 3.4 \\26 & 10 & 10 & 0.077 $\pm$ 0.047 & 25.3 & 0.167 $\pm$ 0.004 & 19.3 & 0.121 $\pm$ 0.011 & 9.2 & 0.153 $\pm$ 0.023 & 7.4 \\27 & -10 & 10 & 0.062 $\pm$ 0.117 & 30.6 & 0.141 $\pm$ 0.005 & 26.4 & 0.104 $\pm$ 0.027 & 12.4 & 0.139 $\pm$ 0.014 & 9.5 \\28 & -30 & 10 & 0.051 $\pm$ 0.044 & 25.1 & 0.089 $\pm$ 0.009 & 21.1 & 0.099 $\pm$ 0.096 & 12.4 & 0.069 $\pm$ 0.015 & 8.8 \\29 & -50 & 10 & -0.032 $\pm$ 0.024 & 10.1 & 0.036 $\pm$ 0.019 & 7.0 & 0.048 $\pm$ 0.031 & 3.4 & 0.126 $\pm$ 0.019 & 5.6 \\30 & 60 & 0 & -0.180 $\pm$ 0.040 & 4.1 & 0.077 $\pm$ 0.755 & 3.5 & 0.002 $\pm$ 0.040 & 2.6 & 0.170 $\pm$ 0.013 & 3.5 \\31 & 40 & 0 & 0.043 $\pm$ 0.064 & 8.9 & 0.060 $\pm$ 0.021 & 10.7 & 0.065 $\pm$ 0.010 & 5.2 & 0.059 $\pm$ 0.019 & 7.1 \\32 & 20 & 0 & 0.083 $\pm$ 0.005 & 58.3 & 0.161 $\pm$ 0.002 & 42.1 & 0.117 $\pm$ 0.010 & 17.5 & 0.032 $\pm$ 0.015 & 7.1 \\33 & 0 & 0 & 0.076 $\pm$ 0.010 & 107.1 & 0.154 $\pm$ 0.001 & 67.1 & 0.106 $\pm$ 0.016 & 40.7 & 0.202 $\pm$ 0.003 & 24.5 \\34 & -20 & 0 & 0.055 $\pm$ 0.023 & 45.7 & 0.134 $\pm$ 0.002 & 31.3 & 0.090 $\pm$ 0.030 & 15.7 & 0.029 $\pm$ 0.013 & 8.0 \\35 & -40 & 0 & 0.046 $\pm$ 0.161 & 13.4 & 0.122 $\pm$ 0.004 & 12.9 & 0.064 $\pm$ 0.052 & 8.3 & 0.055 $\pm$ 0.010 & 9.9 \\36 & -60 & 0 & 0.086 $\pm$ 0.105 & 3.8 & 0.003 $\pm$ 0.008 & 3.9 & -0.065 $\pm$ 0.000 & 3.1 & 0.120 $\pm$ 0.096 & 2.1 \\37 & -80 & 0 & 0.066 $\pm$ 0.000 & 3.2 & -0.151 $\pm$ 0.094 & 2.9 & 0.130 $\pm$ 0.075 & 2.7 & -0.009 $\pm$ 0.021 & 2.5 \\38 & 30 & -10 & 0.020 $\pm$ 0.036 & 17.1 & 0.045 $\pm$ 0.006 & 15.3 & 0.108 $\pm$ 0.008 & 8.7 & 0.106 $\pm$ 0.009 & 7.4 \\39 & 10 & -10 & 0.081 $\pm$ 0.021 & 26.9 & 0.155 $\pm$ 0.003 & 25.6 & 0.100 $\pm$ 0.080 & 12.8 & 0.085 $\pm$ 0.018 & 8.9 \\40 & -10 & -10 & 0.069 $\pm$ 0.031 & 24.9 & 0.081 $\pm$ 0.005 & 24.7 & 0.112 $\pm$ 0.020 & 12.4 & 0.079 $\pm$ 0.013 & 9.1 \\41 & -30 & -10 & 0.043 $\pm$ 0.013 & 22.1 & 0.135 $\pm$ 0.003 & 17.7 & 0.095 $\pm$ 0.005 & 8.9 & 0.040 $\pm$ 0.012 & 8.3 \\42 & -50 & -10 & 0.012 $\pm$ 0.013 & 11.6 & 0.087 $\pm$ 0.006 & 8.4 & 0.101 $\pm$ 0.018 & 4.5 & 0.173 $\pm$ 0.017 & 4.4 \\43 & 80 & -20 & 0.051 $\pm$ 0.008 & 3.4 & 0.137 $\pm$ 0.090 & 3.3 & 0.133 $\pm$ 0.058 & 2.5 & 0.001 $\pm$ 0.018 & 2.1 \\44 & 60 & -20 & 0.001 $\pm$ 0.013 & 5.2 & 0.095 $\pm$ 0.010 & 4.3 & 0.028 $\pm$ 8.160 & 2.9 & 0.059 $\pm$ 0.036 & 2.9 \\45 & 40 & -20 & -0.021 $\pm$ 0.075 & 14.7 & 0.126 $\pm$ 0.004 & 13.6 & 0.114 $\pm$ 0.014 & 7.0 & 0.079 $\pm$ 0.027 & 6.7 \\46 & 20 & -20 & 0.074 $\pm$ 0.043 & 18.8 & 0.075 $\pm$ 0.006 & 17.9 & 0.019 $\pm$ 0.010 & 8.7 & 0.115 $\pm$ 0.015 & 8.5 \\47 & 0 & -20 & 0.072 $\pm$ 0.019 & 49.4 & 0.098 $\pm$ 0.005 & 32.3 & 0.076 $\pm$ 0.058 & 19.8 & 0.045 $\pm$ 0.014 & 7.8 \\48 & -20 & -20 & 0.052 $\pm$ 0.015 & 16.5 & 0.065 $\pm$ 0.008 & 14.7 & 0.095 $\pm$ 0.047 & 8.8 & 0.062 $\pm$ 0.015 & 8.1 \\49 & -40 & -20 & 0.027 $\pm$ 0.011 & 9.7 & 0.072 $\pm$ 0.045 & 10.4 & 0.108 $\pm$ 0.021 & 5.5 & 0.067 $\pm$ 0.018 & 8.8 \\50 & -60 & -20 & 0.004 $\pm$ 1.594 & 3.8 & 0.025 $\pm$ 0.903 & 4.0 & 0.069 $\pm$ 0.029 & 3.0 & -0.015 $\pm$ 0.000 & 3.2 \\51 & 10 & -30 & 0.066 $\pm$ 0.036 & 24.8 & 0.135 $\pm$ 0.004 & 20.3 & 0.176 $\pm$ 0.012 & 9.2 & 0.210 $\pm$ 0.009 & 8.6 \\52 & -10 & -30 & 0.062 $\pm$ 0.038 & 20.8 & 0.107 $\pm$ 0.010 & 19.2 & 0.154 $\pm$ 0.006 & 11.5 & 0.201 $\pm$ 0.009 & 7.0 \\53 & 80 & -40 & 0.064 $\pm$ 0.007 & 4.2 & 0.113 $\pm$ 0.014 & 5.7 & -0.017 $\pm$ 0.048 & 3.1 & 0.143 $\pm$ 0.017 & 4.1 \\54 & 60 & -40 & 0.006 $\pm$ 0.044 & 8.0 & 0.073 $\pm$ 0.465 & 6.9 & 0.016 $\pm$ 0.021 & 3.7 & 0.032 $\pm$ 0.020 & 3.8 \\55 & 40 & -40 & 0.035 $\pm$ 0.018 & 10.8 & 0.145 $\pm$ 0.005 & 8.7 & 0.111 $\pm$ 0.014 & 3.9 & 0.220 $\pm$ 0.013 & 4.2 \\56 & 20 & -40 & 0.062 $\pm$ 0.141 & 13.5 & 0.051 $\pm$ 0.006 & 16.2 & 0.125 $\pm$ 0.007 & 6.8 & 0.198 $\pm$ 0.009 & 8.0 \\57 & 0 & -40 & 0.023 $\pm$ 0.274 & 23.2 & 0.118 $\pm$ 0.003 & 23.7 & 0.164 $\pm$ 0.006 & 13.7 & 0.061 $\pm$ 0.008 & 14.3 \\58 & -20 & -40 & 0.053 $\pm$ 0.168 & 8.0 & 0.060 $\pm$ 0.107 & 10.0 & 0.121 $\pm$ 0.014 & 5.6 & 0.033 $\pm$ 0.016 & 5.7 \\59 & -40 & -40 & 0.091 $\pm$ 0.047 & 4.6 & -0.107 $\pm$ 0.057 & 3.2 & 0.108 $\pm$ 0.039 & 2.6 & 0.054 $\pm$ 0.048 & 2.6 \\60 & 100 & -40 & -0.036 $\pm$ 0.047 & 3.6 & 0.010 $\pm$ 0.018 & 4.5 & 0.106 $\pm$ 0.044 & 2.6 & 0.046 $\pm$ 0.025 & 4.4 \\61 & 100 & -60 & 0.084 $\pm$ 0.048 & 3.2 & 0.048 $\pm$ 0.000 & 3.3 & 0.088 $\pm$ 0.013 & 2.9 & -0.002 $\pm$ 0.030 & 2.2 \\62 & 80 & -60 & 0.098 $\pm$ 0.017 & 7.1 & 0.059 $\pm$ 0.031 & 5.4 & 0.091 $\pm$ 0.123 & 2.7 & 0.028 $\pm$ 0.018 & 4.2 \\63 & 60 & -60 & 0.043 $\pm$ 0.030 & 8.2 & 0.108 $\pm$ 0.015 & 6.3 & 0.017 $\pm$ 0.035 & 3.1 & 0.015 $\pm$ 0.022 & 2.7 \\64 & 40 & -60 & 0.105 $\pm$ 0.013 & 8.1 & 0.077 $\pm$ 0.144 & 7.5 & 0.151 $\pm$ 0.020 & 4.2 & 0.061 $\pm$ 0.028 & 5.1 \\65 & 20 & -60 & 0.020 $\pm$ 0.058 & 9.6 & 0.119 $\pm$ 0.004 & 10.7 & 0.107 $\pm$ 0.025 & 4.1 & 0.096 $\pm$ 0.016 & 5.6 \\66 & 0 & -60 & 0.049 $\pm$ 0.090 & 8.7 & 0.071 $\pm$ 0.110 & 8.4 & 0.094 $\pm$ 0.026 & 3.8 & 0.099 $\pm$ 0.043 & 4.3 \\67 & -20 & -60 & 0.009 $\pm$ 0.013 & 4.9 & 0.038 $\pm$ 0.006 & 3.4 & 0.013 $\pm$ 0.021 & 3.1 & 0.105 $\pm$ 0.061 & 3.3 \\68 & 80 & -80 & -0.012 $\pm$ 0.019 & 3.9 & -0.074 $\pm$ 0.074 & 3.5 & 0.157 $\pm$ 0.025 & 3.1 & 0.095 $\pm$ 0.077 & 2.8 \\69 & 60 & -80 & 0.152 $\pm$ 0.025 & 5.3 & -0.020 $\pm$ 0.025 & 4.2 & 0.124 $\pm$ 0.345 & 2.9 & 0.267 $\pm$ 0.000 & 3.2 \\70 & 40 & -80 & 0.094 $\pm$ 0.013 & 6.7 & 0.074 $\pm$ 0.112 & 6.1 & 0.146 $\pm$ 0.036 & 3.3 & -0.139 $\pm$ 0.024 & 2.8 \\71 & 20 & -80 & 0.103 $\pm$ 0.013 & 6.1 & 0.011 $\pm$ 0.074 & 5.6 & 0.097 $\pm$ 0.028 & 3.0 & 0.128 $\pm$ 0.025 & 2.0 \\72 & 0 & -80 & 0.080 $\pm$ 0.019 & 4.0 & -0.033 $\pm$ 6.564 & 3.6 & 0.083 $\pm$ 0.083 & 2.9 & 0.273 $\pm$ 0.000 & 2.3 \\
\enddata
\tablecomments{Same format as Table 3.  A zero $\pm$ value indicates a low SNR spectrum for which MPFIT was unable to calculate an uncertainty (these sources were omitted from all analyses presented in this paper).  Map central pointing: 19:41:04.5, +10:57:02 (J2000)}
\end{deluxetable}

\begin{deluxetable}{ccccccccccc}
\tabletypesize{\tiny}
\tablewidth{0pt}
\tablecolumns{11}
\tablecaption{L1521F Infall/Expansion Measurements\label{L1521F}}
\tablehead{\colhead{No.} & \colhead{$\Delta\alpha$} & \colhead{$\Delta\delta$} & \colhead{N$_2$H$^+$(1-0)} & \colhead{N$_2$H$^+$(1-0)} & \colhead{DCO$^+$(2-1)} & \colhead{DCO$^+$(2-1)} & \colhead{DCO$^+$(3-2)}  & \colhead{DCO$^+$(3-2)}  & \colhead{HCO$^+$(3-2)} & \colhead{HCO$^+$(3-2)}\\
 & \arcsec & \arcsec & v$_{in}$ (km~s$^{-1}$) & SNR & v$_{in}$ (km~s$^{-1}$) & SNR & v$_{in}$ (km~s$^{-1}$) & SNR & v$_{in}$ (km~s$^{-1}$) & SNR}
\startdata
1 & 0 & 100 & 0.051 $\pm$ 0.261 & 7.7 & 0.019 $\pm$ 1.542 & 5.8 & 0.057 $\pm$ 0.136 & 3.4 & \nodata & \nodata \\2 & -40 & 100 & 0.029 $\pm$ 1.766 & 8.1 & 0.041 $\pm$ 0.067 & 5.1 & 0.046 $\pm$ 0.218 & 2.0 & \nodata & \nodata \\3 & 40 & 80 & 0.128 $\pm$ 0.011 & 5.4 & 0.101 $\pm$ 0.035 & 2.8 & 0.016 $\pm$ 0.024 & 3.2 & \nodata & \nodata \\4 & 0 & 80 & 0.054 $\pm$ 0.045 & 15.4 & 0.062 $\pm$ 0.020 & 9.3 & -0.006 $\pm$ 0.019 & 6.4 & \nodata & \nodata \\5 & -20 & 80 & 0.065 $\pm$ 0.064 & 14.1 & 0.040 $\pm$ 0.008 & 11.4 & -0.038 $\pm$ 0.040 & 5.5 & \nodata & \nodata \\6 & -40 & 80 & 0.002 $\pm$ 0.008 & 10.0 & -0.052 $\pm$ 0.021 & 7.7 & -0.014 $\pm$ 0.024 & 3.2 & \nodata & \nodata \\7 & -60 & 80 & -0.015 $\pm$ 0.011 & 9.4 & -0.054 $\pm$ 0.026 & 7.1 & -0.011 $\pm$ 1.614 & 4.0 & \nodata & \nodata \\8 & 20 & 60 & 0.023 $\pm$ 0.449 & 16.3 & 0.042 $\pm$ 0.032 & 8.0 & -0.045 $\pm$ 0.029 & 3.6 & \nodata & \nodata \\9 & 0 & 60 & 0.110 $\pm$ 0.005 & 20.0 & 0.106 $\pm$ 0.016 & 16.7 & 0.076 $\pm$ 0.110 & 11.5 & \nodata & \nodata \\10 & -20 & 60 & 0.116 $\pm$ 0.005 & 20.1 & 0.097 $\pm$ 0.022 & 10.9 & 0.029 $\pm$ 0.045 & 6.4 & \nodata & \nodata \\11 & -40 & 60 & 0.079 $\pm$ 0.007 & 18.4 & 0.080 $\pm$ 0.029 & 9.5 & -0.000 $\pm$ 0.090 & 3.8 & \nodata & \nodata \\12 & 60 & 40 & 0.004 $\pm$ 0.892 & 6.4 & 0.026 $\pm$ 0.027 & 2.9 & -0.013 $\pm$ 0.051 & 2.6 & \nodata & \nodata \\13 & 40 & 40 & 0.062 $\pm$ 0.080 & 15.0 & -0.139 $\pm$ 0.019 & 5.1 & 0.001 $\pm$ 4.471 & 2.9 & \nodata & \nodata \\14 & 20 & 40 & 0.080 $\pm$ 0.008 & 18.8 & 0.026 $\pm$ 0.891 & 8.5 & 0.084 $\pm$ 0.031 & 6.5 & \nodata & \nodata \\15 & 0 & 40 & 0.100 $\pm$ 0.021 & 21.3 & 0.117 $\pm$ 0.025 & 9.2 & 0.085 $\pm$ 0.239 & 6.1 & \nodata & \nodata \\16 & -20 & 40 & 0.087 $\pm$ 0.004 & 51.2 & 0.138 $\pm$ 0.004 & 22.0 & 0.044 $\pm$ 0.137 & 15.2 & 0.062 $\pm$ 0.008 & 14.5 \\17 & -40 & 40 & 0.100 $\pm$ 0.025 & 22.0 & 0.165 $\pm$ 0.008 & 7.7 & 0.091 $\pm$ 0.061 & 5.1 & \nodata & \nodata \\18 & -60 & 40 & 0.062 $\pm$ 0.058 & 19.1 & -0.007 $\pm$ 0.022 & 7.9 & -0.012 $\pm$ 0.017 & 4.9 & \nodata & \nodata \\19 & 40 & 20 & -0.076 $\pm$ 0.015 & 13.2 & -0.058 $\pm$ 1.108 & 4.5 & -0.007 $\pm$ 1.595 & 2.2 & \nodata & \nodata \\20 & 20 & 20 & -0.041 $\pm$ 0.827 & 14.0 & -0.008 $\pm$ 0.014 & 7.0 & -0.011 $\pm$ 1.621 & 3.9 & \nodata & \nodata \\21 & 0 & 20 & 0.096 $\pm$ 0.030 & 47.3 & 0.117 $\pm$ 0.034 & 15.5 & -0.070 $\pm$ 0.162 & 12.8 & 0.071 $\pm$ 0.006 & 21.9 \\22 & -20 & 20 & 0.092 $\pm$ 0.004 & 129.2 & 0.107 $\pm$ 0.002 & 66.8 & 0.038 $\pm$ 0.144 & 46.5 & 0.041 $\pm$ 0.006 & 23.0 \\23 & -40 & 20 & 0.080 $\pm$ 0.016 & 28.1 & 0.093 $\pm$ 0.030 & 7.7 & 0.079 $\pm$ 0.024 & 7.6 & -0.020 $\pm$ 0.009 & 9.6 \\24 & -60 & 20 & 0.044 $\pm$ 0.138 & 11.1 & 0.028 $\pm$ 1.658 & 4.9 & -0.018 $\pm$ 0.026 & 3.4 & \nodata & \nodata \\25 & -80 & 20 & -0.010 $\pm$ 1.012 & 9.1 & 0.012 $\pm$ 0.010 & 3.4 & 0.010 $\pm$ 0.206 & 2.4 & \nodata & \nodata \\26 & 60 & 0 & -0.018 $\pm$ 0.607 & 6.4 & -0.128 $\pm$ 0.027 & 2.8 & 0.177 $\pm$ 0.027 & 3.0 & \nodata & \nodata \\27 & 40 & 0 & -0.039 $\pm$ 0.028 & 9.4 & \nodata & \nodata & \nodata & \nodata & -0.052 $\pm$ 0.020 & 3.4 \\28 & 20 & 0 & -0.100 $\pm$ 0.016 & 26.2 & -0.061 $\pm$ 0.017 & 10.4 & 0.023 $\pm$ 1.582 & 5.5 & -0.001 $\pm$ 0.010 & 7.3 \\29 & 0 & 0 & 0.056 $\pm$ 0.004 & 33.7 & 0.099 $\pm$ 0.005 & 27.6 & 0.081 $\pm$ 0.110 & 16.5 & \nodata & \nodata \\30 & -20 & 0 & 0.082 $\pm$ 0.007 & 45.7 & 0.162 $\pm$ 0.004 & 16.3 & 0.076 $\pm$ 0.171 & 9.5 & 0.013 $\pm$ 0.008 & 8.3 \\31 & -40 & 0 & 0.081 $\pm$ 0.063 & 18.6 & 0.076 $\pm$ 0.015 & 8.3 & 0.101 $\pm$ 0.078 & 3.6 & \nodata & \nodata \\32 & -60 & 0 & 0.079 $\pm$ 0.024 & 8.3 & -0.027 $\pm$ 3.740 & 3.4 & -0.052 $\pm$ 0.021 & 3.7 & \nodata & \nodata \\33 & 40 & -20 & -0.053 $\pm$ 0.334 & 7.6 & \nodata & \nodata & \nodata & \nodata & -0.199 $\pm$ 0.122 & 2.6 \\34 & 20 & -20 & -0.032 $\pm$ 0.018 & 10.8 & \nodata & \nodata & \nodata & \nodata & 0.110 $\pm$ 0.075 & 3.3 \\35 & 0 & -20 & 0.084 $\pm$ 0.058 & 23.7 & 0.126 $\pm$ 0.015 & 16.4 & 0.069 $\pm$ 0.301 & 10.2 & \nodata & \nodata \\36 & -20 & -20 & 0.076 $\pm$ 0.050 & 23.0 & 0.093 $\pm$ 0.077 & 10.9 & -0.024 $\pm$ 1.858 & 6.6 & \nodata & \nodata \\37 & -40 & -20 & 0.087 $\pm$ 0.050 & 13.8 & 0.020 $\pm$ 0.032 & 4.7 & 0.043 $\pm$ 0.182 & 4.9 & \nodata & \nodata \\38 & 20 & -40 & 0.062 $\pm$ 0.110 & 6.5 & \nodata & \nodata & \nodata & \nodata & 0.134 $\pm$ 0.090 & 3.0 \\39 & 0 & -40 & 0.063 $\pm$ 0.048 & 11.0 & 0.114 $\pm$ 0.022 & 6.2 & 0.089 $\pm$ 0.013 & 4.2 & \nodata & \nodata \\40 & -20 & -40 & 0.042 $\pm$ 0.018 & 9.7 & 0.074 $\pm$ 0.005 & 6.2 & 0.145 $\pm$ 0.018 & 3.4 & \nodata & \nodata \\41 & -60 & -40 & 0.044 $\pm$ 0.000 & 3.1 & -0.002 $\pm$ 0.008 & 2.7 & 0.046 $\pm$ 0.207 & 2.7 & \nodata & \nodata \\42 & 0 & -60 & 0.047 $\pm$ 0.055 & 5.9 & 0.052 $\pm$ 0.032 & 3.2 & 0.150 $\pm$ 0.000 & 2.5 & \nodata & \nodata \\
\enddata
\tablecomments{Same format as Table 3.  Blank entries indicate the position was not observed using the given column's tracer. A zero $\pm$ value indicates a low SNR spectrum for which MPFIT was unable to calculate an uncertainty (these sources were omitted from all analyses presented in this paper).  Map central pointing: 04:28:39.8, +26:51:15 (J2000)}

\end{deluxetable}

\begin{deluxetable}{ccccccc}
\tabletypesize{\tiny}
\tablewidth{0pt}
\tablecolumns{7}
\tablecaption{Fraction of SNR$>$6 spectra better fit by HILL5 model according to F-test \label{F-test}}
\tablehead{\colhead{Core} & \colhead{N$_2$H$^+$(1-0)} & \colhead{DCO$^+$(2-1)} & \colhead{DCO$^+$(3-2)} & \colhead{HCO$^+$(3-2)} }
\startdata
L492 & 59$\%$ & 44$\%$ & 0$\%$\tablenotemark{a} & 0$\%$\tablenotemark{a} \\
L694-2 & 80$\%$ & 86$\%$ & 67$\%$ & 95$\%$ \\ 
L1521F & 72$\%$ & 88$\%$ & 46$\%$ & 100$\%$ \\ 
\enddata
\tablenotetext{a}{Single pointing}
\tablecomments{Percentages represent the fraction of spectra that reject our null hypothesis (see section 3.3) at the 90$\%$ confidence level (significance of $\alpha$ = 0.10).}
\end{deluxetable}

\begin{deluxetable}{ccccc}
\tabletypesize{\tiny}
\tablewidth{0pt}
\tablecolumns{5}
\tablecaption{Median Infall Velocity Measurements\label{Average}}
\tablehead{\colhead{Core} & \colhead{N$_2$H$^+$(1-0)} & \colhead{DCO$^+$(2-1)} & \colhead{DCO$^+$(3-2)} & \colhead{HCO$^+$(3-2)}\\
 & km~s$^{-1}$ & km~s$^{-1}$ & km~s$^{-1}$ & km~s$^{-1}$}
\startdata
L492 & 0.032 $\pm$ 0.008 & 0.066 $\pm$ 0.017 & 0.055\tablenotemark{a} $\pm$ 0.137 & 0.029\tablenotemark{a} $\pm$ 0.500\\ 
L694-2 & 0.052 $\pm$ 0.005 & 0.105 $\pm$ 0.005 & 0.100 $\pm$ 0.009 & 0.106 $\pm$ 0.011 \\ 
L1521F & 0.062 $\pm$ 0.009 & 0.093 $\pm$ 0.013 & 0.069 $\pm$ 0.013 & 0.027 $\pm$ 0.013 \\ 
\enddata
\tablenotetext{a}{Single pointing; uncertainty from individual HILL5 model fit to spectrum (see text).}
\tablecomments{Calculated using spectra with SNR $>$ 6. All medians contain 5 or more pointings and uncertainties are calculated using $\sigma$/$\sqrt{N}$, unless otherwise noted.}
\end{deluxetable}

\end{document}